\documentclass[10pt,aps,prc,twocolumn,showpacs,showkeys,amsmath,floatfix,superscriptaddress,preprintnumbers]{revtex4-1}
\usepackage{color,graphicx}
\usepackage{mathptmx}                
\usepackage{dcolumn}                 
\usepackage{bm}                      
\usepackage{ulem,soul}
\usepackage{nicefrac}
\usepackage{multirow}
\usepackage[dvipsnames,usenames]{xcolor}
\DeclareMathAlphabet{\mathpzc}{OT1}{pzc}{m}{it}

\begin{document}


\title{Bayesian analysis of the crust-core transition with a compressible liquid-drop model}

\author{Thomas Carreau}
\author{Francesca Gulminelli}
\affiliation{CNRS, ENSICAEN, UMR6534, LPC ,F-14050 Caen Cedex, France}

\author{J\'er\^ome Margueron}
\affiliation{Institut de Physique Nucl\'eaire de Lyon, CNRS/IN2P3, Universit\'e de Lyon, Universit\'e Claude Bernard Lyon 1, F-69622 Villeurbanne Cedex, France}

\date{\today}

\begin{abstract}
The crust-core phase transition of neutron stars is quantitatively studied within a unified meta-modelling of the nuclear Equation of State (EoS). The variational equations in the crust are solved within a Compressible Liquid Drop (CLD) approach, with surface parameters consistently optimized for each EoS set on experimental nuclear mass data. 
When EoS parameters are taken from known Skyrme or RMF functionals, the transition point of those models is nicely reproduced.
A model-independent probability distribution of EoS parameters and of the transition density and pressure is determined with a Bayesian analysis, where the prior is given by an uncorrelated distribution of parameters within the present empirical uncertainties, and constraints are applied both from neutron star physics and ab-initio modelling. We show that the characteristics of the transition point are largely independent of the high density properties of the EoS, while ab-initio EoS calculations of neutron and symmetric matter are far more constraining. The most influential parameter for the determination of the transition point governs the surface properties of extremely neutron rich matter, and it is strongly unconstrained. This explains the large dispersion of existing predictions of the transition point. Only if the surface tension is fixed to a reasonable but somewhat arbitrary value, strong correlations with isovector EoS parameters ($L_{sym},K_{sym}$ and $Q_{sym}$) are recovered. Within the present experimental and theoretical uncertainties on those parameters, we estimate the transition density as $n_t= 0.072\pm 0.011$ fm$^{-3}$ and the transition pressure as $P_t=0.339\pm0.115$ MeV fm$^{-3}$. 
\end{abstract}

\maketitle

\section{Introduction}

Neutron stars (NS) are a unique observable laboratory of the different phases of hadronic matter~\cite{Haensel_book}.
The structure and composition of the inner core is still not completely clear~\cite{report_Micaela}, but it is however well established that a phase transition occurs from a solid crust to a liquid core at some $\approx 1$ km from the surface of the star. 
The presence of a solid crust plays an important role in a number of phenomena involving NS, from the cooling of proto-neutron stars~\cite{Page2013,Shternin2007,Brown2009} to the irregularities ("glitches") in their rotational motion~\cite{Haskell2015,Link1999,Steiner2015}. The precise location of the transition in the star is also important for the theoretical determination of its static properties, notably the star radius~\cite{Fortin2016}, which accurate measurement will be soon available with upcoming x-ray observations~\cite{Watts2016}. To understand all these aspects of NS physics, a reliable theoretical estimation of the crustal thickness and its uncertainty is necessary. 

For slowly rotating NS, the crustal thickness can be computed with the Tolman-Oppenheimer-Volkov (TOV) equation of hydrostatic equilibrium, if the equation of state (EoS) and the core-crust (CC) transition point  is known in density $n_t$ and pressure $P_t$. 
A large number  of studies have been devoted to the determination of the transition point with different relativistic~\cite{Horowitz2001,Klahn2006,Moustakidis2010,Fattoyev2010,Cai2012,Pais2016} and non relativistic~\cite{Vidana2009,Li2016,Ducoin2011,Routray2016,Gonzales2017} models. Most of these studies compute the transition "from the core": the transition is defined as the density point where homogeneous nuclear matter becomes unstable with respect to density fluctuations. The simplest version of this technique consists in evaluating the thermodynamical spinodal, that is the instability point of neutral nuclear matter with respect to the nuclear liquid-gas (LG) phase transition. The advantage of this method is that it only requires the knowledge of the energy functional of homogeneous nuclear matter: this functional, at the density of the CC point which is close to nuclear saturation density, is characterized by a small set of parameters which can be strongly constrained through nuclear experiments and/or microscopic ab-initio calculations~\cite{Tsang2012,Lattimer2013,Dutra2014,Fortin2016}. 

However, the CC transition is very different from the LG one~\cite{Ducoin2007,Chomaz2007} and the thermodynamical spinodal gives only a qualitative (and overestimated) estimation of the transition point. The CC transition occurs at the pressure where the  energy density of clusterized matter (the solid cristal embedded in a electron and neutron gas) overcomes the energy density of uniform nuclear matter~\cite{Baym1971}. The clusterized phase is inhomogeneous and locally charged, and its equilibrium energy is determined by the competition between the Coulomb and the surface energy. These energy components identically vanish in uniform matter, and therefore do not enter in the determination of the thermodynamical spinodal. 

A better estimation can be obtained looking at the dynamical response of the homogeneous system with respect to finite size fluctuations~\cite{Pethick1995}. The resulting dynamical spinodal is available for a limited set of models~\cite{Ducoin2007,Pais2016,Xu2009,Pearson2012}, and was recently studied by our group within a Bayesian meta-modelling technique~\cite{Chatterjee,Antic}. 
One should remark that  for the determination of the dynamical spinodal, in addition to the EoS of uniform matter, the isovector gradient terms - which are not well known - play a non negligible role,
and its  precise location slightly depends on the many body formalism adopted (linear response, Vlasov, or RPA)~\cite{Ducoin2008}.  
 Even if the dynamical spinodal certainly gives a better estimation of the transition than the thermodynamical one, it is worth mentioning that in the actual dynamical process of supernova collapse that gives birth to neutron stars, matter at subsaturation densities is never uniform but composed of atomic nuclei~\cite{Lattimer1991,Shen1998}, meaning that spinodal decomposition is not the dynamical process leading to the formation of the crust. 

For this reason, the most  theoretically sound determination of the CC point consists in determining the transition "from the crust", by directly comparing the energy density of the two competing phases~\cite{Baym1971}. This method demands an explicit modelling of clusterized matter in beta equilibrium, which is a complex quantum many body problem. Therefore, after the seminal work by Baym, Bethe and Pethick~\cite{Baym1971}, only  few works using modern energy functionals have been developed along this line~\cite{Avancini2008,Goriely2010,Sharma2015,Douchin2001,Newton2013,Gulminelli2015},  and the problem of model dependence clearly arises. In particular, many works have been devoted to the EoS dependence of the transition point and in particular the effect of the $L_{sym}$ parameter~\cite{Pearson2012}. However, the interplay between the EoS parameters and the the isovector surface tension coefficient, which from the microscopic viewpoint is determined by the isovector gradient terms in the energy functional mentioned above, has been seldom addressed~\cite{Newton2013}.

In this paper, we will calculate the CC transition point "from the crust", by solving the variational equations for non-uniform matter within a compressible liquid drop (CLD) approach, in the same lines of Refs.~\cite{Baym1971,Douchin2001,Gulminelli2015}. 
The distribution probability of the EoS parameters of uniform matter and the extra parameters associated to the cluster surface properties will be determined with a Bayesian analysis, using a fully uncorrelated flat prior and constraints from the low density effective field theory (EFT) modelling by Drischler et al.~\cite{Drischler2016}. This will allow us presenting model independent estimations of the density and pressure of the transition point, and determining which are the most influential parameters governing the phase transition.

The theoretical uncertainties on the EoS and on the transition point propagate to global observabes of the neutron star such as the crust thickness and moment of inertia, which in turn  may be linked to astrophysical observables such as the amplitude of 
neutron star glitches. Quantative predictions on these global observables  were presented in a recent paper~\cite{Thomas_prl}. In the present  work, we concentrate on the nuclear physics ingredients, namely the density and pressure of the transition point,
and examine in greater details the influence of the different parameters, and the effectiveness of the different constraints.  

A very similar Bayesian  study was very recently and independently performed in Ref.~\cite{Holt}, for the computation of different quantities than the ones of the present work, namely radii and tidal polarizabilities. 
The functional expression chosen for the homogeneous matter EoS is not the same as in our work, but 
 our posterior distributions for the EOS parameters are in very good agreement with the results of Ref.~\cite{Holt}, showing the reliability and generality of the meta-modelling technique.

The plan of the paper is as follows. In section~\ref{sec:modelling} we describe the variational equations which are solved to determine the crust composition and the transition to the core. The meta-modelling technique from Ref.~\cite{Margueron2018a,Margueron2018b} used for the uniform matter EoS and the expression of the cluster surface tension from Ref.~\cite{Ravenhall1983,Lorentz1993} will also be shortly summarized. Section~\ref{sec:results} demonstrates the ability of our meta-modelling technique to reproduce the published results of specific models. The crust composition and the CC point obtained with specific choices for the model parameters will be compared to the litterature, showing that the parameter space of our meta-modelling is large enough to cover existing functionals, and can thus be used for a Bayesian determination of the EoS parameters.
Section~\ref{sec:sensitivity} presents a sensitivity analysis to the different EoS parameters, and we will show that, together with the slope of the symmetry energy $L_{sym}$, the curvature $K_{sym}$ is strongly influential in the determination of the CC pressure.
The full Bayesian analysis is reported in section~\ref{sec:bayes}, where the probability distribution for the transition observables is computed imposing to our uncorrelated prior to reproduce the band predictions in isospin-symmetric and neutron matter
 of Ref.~\cite{Drischler2016} obtained from a many-body perturbation theory (MBPT) based on two and three-nucleon chiral EFT interactions at N$^3$LO. We will show that these ab-initio calculations at low density are far more constraining than the astrophysical constraint at high density concerning the maximum mass of NS. A complete correlation study will also be presented, where the importance of the isovector surface energy will be underlined. Conclusions are drawn in section~\ref{sec:conclusions}.



\section{Modelling inhomogeneous matter}\label{sec:modelling}

\subsection{Variational equations}
The equilibrium configuration of inhomogeneous catalyzed matter is obtained following the standard variational formalism of Refs.~\cite{Baym1971,Douchin2001,Gulminelli2015}. Using the Lagrange multipliers technique, the energy density in a Wigner-Seitz cell of volume $V_{WS}$ is minimized with the constraint of a given baryonic density $n_B=n_p+n_n$. The auxiliary function to be minimized reads:
\begin{equation}
\mathpzc{F}(A,I,n_0,n_p,n_g)=\frac{E_{nuc}}{V_{WS}}+\left (1-\frac{A}{n_0V_{WS}}\right )\epsilon_g+\epsilon_{el}-\mu n_B, \label{eq:auxiliary}
\end{equation}
where $\epsilon_g=\epsilon(n_p=0,n_n=n_g)$ ($\epsilon_{el}$) is the energy density of a pure uniform neutron (electron) gas at density $n_g$ ($n_e$), and the bulk interaction between the cluster and the neutron gas is treated in the excluded volume approximation. The cluster energy $E_{nuc}$ depends on the cluster atomic number $A$, isospin asymmetry $I=(N-Z)/A$ and density $n_0$, and also on the total electron density $n_e=n_p$ because of the electrostatic interaction with the electron gas, according to:
\begin{equation}
E_{nuc}=\frac{\epsilon(n_{0p},n_{0n})}{n_0}A+E_c+E_s, \label{eq:enuc}
\end{equation}
where $n_{0n(p)}=n_0(1\pm I/2)$, $E_s$ is the surface energy to be discussed in section~\ref{sec:surface} below, and we use the standard expression for the Coulomb energy $E_c$ from Ref.~\cite{Baym1971,Gulminelli2015}. 
Minimizing with respect to the five independent variables $A,I,n_0,n_p,n_g$, and using the baryonic density constraint,
\begin{equation}
n_B=n_g+\frac{A}{V_{WS}}\left (1-\frac{n_g}{n_0} \right ),
\end{equation}
leads to the following system of coupled differential equations:

\begin{eqnarray}
    \frac{\partial (E_{nuc}/A)}{\partial A}\bigg|_{I,n_0,n_p,n_g} = 0, \label{eq:eq1} \\
    \frac{2}{A}\frac{\partial E_{nuc}}{\partial I}\bigg|_{A,n_0,n_p,n_g} = \mu_{el} - n_p\frac{\partial (E_c/A)}{\partial n_p}\bigg|_{A,I,n_0}, \label{eq:eq2} \\
    \frac{E_{nuc}}{A} + \frac{1-I}{A}\frac{\partial E^{nuc}}{\partial I}\bigg|_{A,n_0,n_p,n_g} - \frac{\epsilon_g}{n_0} = \mu\left(1-\frac{n_g}{n_0}\right), \label{eq:eq3} \\
    {n_0}^2\frac{\partial (E_{nuc}/A)}{\partial n_0}\bigg|_{A,I,n_p,n_g} = n_g\mu - \epsilon_g, \label{eq:eq4}
\end{eqnarray}

where the baryonic chemical potential $\mu$ results:

\begin{equation}
    \mu = \frac{2 n_p}{n_0A(1-I) - 2 n_p}\frac{\partial E_s}{\partial n_g}\bigg|_{A,I,n_0} + \frac{d\epsilon_g}{dn_g} .
\end{equation}

We can see that, in the absence of a possible in-medium modification of the surface energy because of the external gas, the baryonic chemical potential
can be identified with the chemical potential of the gas $\mu_g\equiv d \epsilon_g/ d n_g$.

It  is easy to show that equations (\ref{eq:eq1})-(\ref{eq:eq4}) can be equivalently written as chemical and mechanical equilibrium equations between the cluster and the neutron and electron gas, supplemented by the Baym virial theorem~\cite{Baym1971}:
\begin{eqnarray}
\mu_n^{nuc}&=&\mu_p^{nuc}+\mu_{el}+\Delta \mu, \\
\mu_n^{nuc}&=&\mu_g, \\
P_{nuc}&=&P_g, \\
E_s&=&2E_c,
\end{eqnarray}
where $P_g=n_g\mu_g-\epsilon_g$, the cluster chemical potential is modified by the external neutron gas as:
\begin{equation}
\mu_n^{nuc}=\frac{\partial E^{nuc}}{\partial N}\bigg|_Z+\frac{P_g}{n_0},
\end{equation}
and the electrostatic interaction between protons in the cluster and the background electrons leads to a modification of the $\beta$-equilibrium condition, $\Delta \mu=-n_p\partial (E_c/A)/\partial n_p$.

Equations (\ref{eq:eq1})-(\ref{eq:eq4}) can be numerically solved if the energy functional for homogeneous baryonic matter $\epsilon(n_p,n_n)$ and the surface energy $E_s$ are specified. These quantities are affected by strong uncertainties, especially in the isovector sector~\cite{report_Micaela,Steiner2013,Newton2013,Newton2014}. To quantify the uncertainty on the crust-core transition induced by our imperfect knowledge of the baryonic bulk and surface energy, we use for both quantities simple and flexible parametrized expressions with parameters whose variation embeds the present uncertainty on nuclear energetics. 
These functionals are presented in sections~\ref{sec:metamodel} and~\ref{sec:surface} below.

\subsection{EoS meta-modelling}\label{sec:metamodel}

We parametrize the energy density of homogeneous nuclear matter with baryonic density $n=n_n+n_p$ and isospin asymmetry $\delta=(n_n-n_p)/n$,  using the meta-modelling technique of Refs.~\cite{Margueron2018a,Margueron2018b}, here briefly summarized. 

It is theoretically known that the energy density of homogeneous nuclear matter is analytic at least up to $\approx 2-3 n_{sat}$, where $n_{sat}$ is the saturation density of symmetric matter. The integrality of  the possible behaviors of the functional can therefore be explored using a Taylor expansion around the saturation point $(n=n_{sat},\delta=0)$, and largely varying the parameters of the expansion, that correspond to the well-known EoS empirical parameters~\cite{Piekarewicz2009}:
\begin{eqnarray}
\epsilon(n,\delta)&=& \sum_{m\ge 0} \frac{1}{m!} \sum_{k\ge 0} C_m^k \delta^k x^m ,
\label{eq:etotisiv}
\end{eqnarray} 
with $x=(n-n_{sat})/3n_{sat}$. To fasten the series convergence, the $\delta^{5/3}$ term coming from the fermionic zero point energy is explicitly added, as well as an exponential correction insuring the correct limiting behavior at zero density, see Ref.~\cite{Margueron2018a} for more details.

The final form for the energy per particle $e(n,\delta)=\epsilon/n$ at order $N$ of the expansion is given by:
\begin{eqnarray}
e(n,\delta)&=&  \frac{3\hbar^{2}}{20m}\left(\frac{3\pi^{2}n}{2}\right)^{2/3}
\bigg[ \left( 1+\kappa_{sat}\frac{n}{n_{sat}} \right) f_1
+ \kappa_{sym}\frac{n}{n_{sat}}f_2\bigg] \nonumber  \\
&+&\sum_{m\geq0}^N ( v_{m}^{is}+ v_{m}^{iv} \delta^2) \frac{x^m}{m!}
-(a_N^{is}+a_N^{iv}\delta^2) x^{N+1} e^{ -b\frac{n}{n_{sat}}},
\label{eq:vELFc}
\end{eqnarray}
where the  functions $f_1, f_2$ give an effective correction to the parabolic approximation for the symmetry energy:
\begin{eqnarray}
f_1(\delta) &=& (1+\delta)^{5/3}+(1-\delta)^{5/3} \\
f_2(\delta) &=& \delta \left( (1+\delta)^{5/3}-(1-\delta)^{5/3} \right) .
\end{eqnarray}
The parameters $v_m^{is}, v_m^{iv}, a_N^{is}, a_N^{iv}$ of the meta-functional eq.(\ref{eq:vELFc}) are linear combinations of the  successive derivatives $E_q,L_q, K_q, Q_q, Z_q,\dots$ in the isoscalar ($q=sat$) and isovector  ($q=sym$) sector. The saturation density $n_{sat}$, two parameters linked to the isoscalar effective mass ($\kappa_{sat}$) and effective mass splitting ($\kappa_{sym}$), and the $b$ parameter governing the functional behavior close to the $n\to 0$
limit~\cite{Margueron2018a}, complete the parameter set, which we will note in a compact form as $\vec X\equiv\{X_n, n=1,\dots, 2(N+1)+3\}$. It was verified in Ref.~\cite{Margueron2018a} that with an expansion up to $N=4$, the parameter space  is sufficiently large to give an excellent reproduction of different popular Skyrme and RMF functionals. By varying the values of $\vec X$, eq.(\ref{eq:vELFc}) thus provides a meta-functional that can continuously interpolate among existing  functionals, and possibly explore novel density dependencies which have not yet been proposed in the literature.

A flat distribution of $\vec X$ values within physically reasonable intervals determined by empirical evidence~\cite{Margueron2018a}, will give our prior distribution for the EoS parameters. Ab-initio MBPT calculations of uniform nuclear matter~\cite{Drischler2016} will be used in section~\ref{sec:bayes} as a constraint, to determine a posterior distribution that will be used to estimate the model dependence of the crust-core transition calculation. A similar strategy was employed  to compute
different observables of neutron stars in Ref.~\cite{Margueron2018b,Thomas_prl,Chatterjee,Antic,Holt}, and of finite nuclei in Ref.~\cite{Chatterjee2017,Holt}.

\subsection{Surface tension}\label{sec:surface}

To determine the crust-core phase transition, the energy density of homogeneous matter must be compared with the energy density of clusterized matter. The difference between the two is essentially given by the Coulomb $E_c$ and surface $E_s$ contribution (see eq.(\ref{eq:enuc})), which identically vanish in homogeneous matter. The Coulomb energy only depends on the cluster charge $Z$ and density $n_0$, which are consistently obtained from the variational equations (\ref{eq:eq1})-(\ref{eq:eq4}).
On the other hand, to calculate the surface energy from the energy functional, strong approximations are needed in the many-body treatment~\cite{Baym1971,Ravenhall1972}. Within these approximations, gradient terms must be specified, involving extra parameters which are added in the fitting protocol of phenomenological functionals~\cite{Bender2011}. General physical considerations indicate that the surface tension $\sigma(A,I)=E_s/S$, with $S$ nuclear surface, should strongly depend on the
isospin asymmetry $I$ (with $\sigma\to 0$ as $I\to 1$)~\cite{Centelles1998,Danielewicz2003} and only moderately on the nuclear mass $A$~\cite{Durand1993},  but the exact value of $\sigma(I)$ is model dependent, and the uncertainty is particularly important for the extreme $I$ values encountered in the inner crust: indeed no experimental data exist on the surface energy of nuclei beyond the drip-line, which is in-medium modified by the presence of the neutron gas~\cite{Centelles1998,Douchin2000,Sil2002,Aymard2014}. For this reason, parametrized expressions are generally employed~\cite{Steiner2005}.

\begin{figure}[htbp]
    \centering
    \includegraphics[scale=0.35]{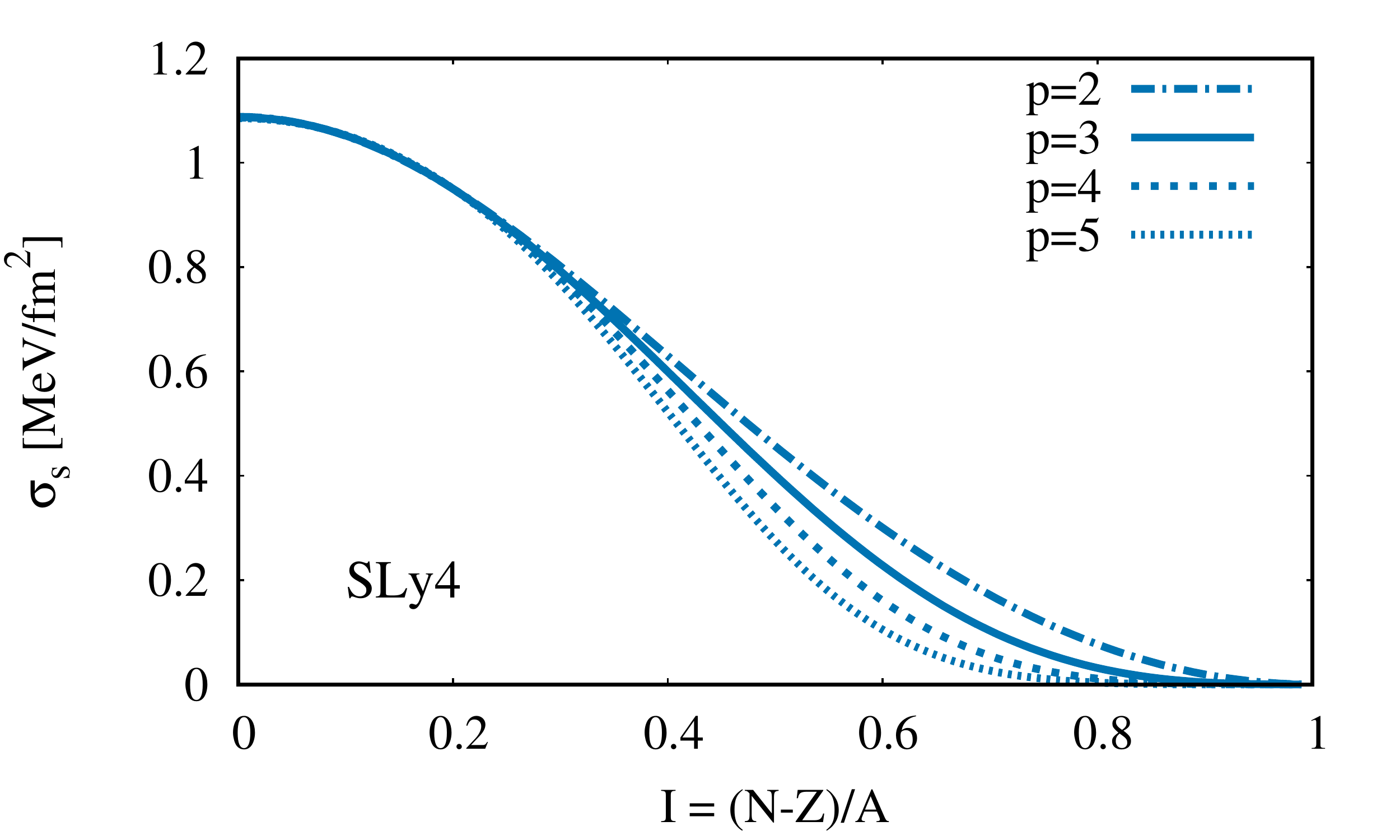}
    \caption{Surface tension eq.(\ref{eq:sigma}) as a function of the isospin asymmetry for different values of the $p$ parameter for the SLy4 functional.} 
 \label{fig:surface_tension}
\end{figure}

We use the expression originally proposed by Ravenhall et al.~\cite{Ravenhall1983} on the basis of Thomas-Fermi calculations at extreme isospin ratios, and later employed in different works on neutron star and supernova modelling within the compressible liquid-drop model~\cite{Lorentz1993,Lattimer1991,Newton2013}:
\begin{equation}
  E_s(A,Y_p) = 4\pi r_{sat}^2A^{2/3}\sigma(Y_p),
\end{equation}

with $r_{sat}=(3/4\pi n_{sat})^{1/3}$, $Y_p=Z/A$, and

\begin{equation}
    \sigma(Y_p) = \sigma_0\frac{2^{p+1}+b_s}{Y_p^{-p}+b_s+(1-Y_p)^{-p}}. \label{eq:sigma}
\end{equation}

In this expression, the parameter $\sigma_0$ determines the value of the surface tension of symmetric nuclei, while $b_s$ governs the isospin dependence for moderate asymmetries. These  parameters are fitted from experimental masses of the spherical magic and semi-magic nuclei $^{40,48}$Ca, $^{48,58}$Ni, $^{88}$Sr, $^{90}$Zr, $^{114,132}$Sn, and $^{208}$Pb. Enlarging the set of mass data does not modify the results presented in this paper. This fit provides optimal values for $\sigma_0$ and $b_s$ different for each set of uniform matter parameters $\vec X$. For illustration, Figure~\ref{fig:surface_tension} displays the surface tension when $\sigma_0$ and $b_s$ are fixed for the parameter set $\vec X$ corresponding to the SLy4 interaction~\cite{Chabanat1997}. Different values for the parameter $p$ are also displayed in that figure. We can see that the $p$ parameter determines the behavior of the surface tension for extreme isospin values~\cite{Newton2013}, and cannot be accessed from empirical nuclear physics data, which are limited to values around $I\leq 0.3$.
For this reason, the parameter $p$ will be added to the $\vec X$ set as an extra dimension in the parameter space of our meta-modelling, and a reasonable variation interval for its prior distribution will be determined in section~\ref{sec:sensitivity}.

\section{Results for a representative EoS}\label{sec:results}

In the previous section we have seen that, even in a unified equation of state approach as the one employed in this paper, many different parameters have to be specified to calculate the crust composition and the crust-core phase transition. The importance of the equation of state empirical parameters like $E_{sym}, L_{sym}$ has been pointed out by many authors in the context of specific EoS models~\cite{Ducoin2011} and their influence will be studied in section~\ref{sec:sensitivity}. In addition to that, other parameters, much less constrained in the neutron star and neutron physics literature, enter explicitly in the variational equations eqs.(\ref{eq:eq1})-(\ref{eq:eq4}). These parameters comprise the high order derivatives $Q_q,Z_q,\dots$ ($q=sat,sym$), the $b$ parameter that determines the limiting low density value where the Taylor expansion around saturation breaks down (see eq.(\ref{eq:vELFc})), and the $p$ parameter discussed in section~\ref{sec:surface} governing the isovector behavior of the surface energy at extreme isospin values. In this section we will therefore consider a representative EoS model, and investigate whether the transition observables are modified by varying these relatively uncontrolled parameters. The representative EoS model is taken to be the parameter set $\vec X$ with empirical parameters fixed from the SLy4 functional, and will be noted as meta-SLy4 in the following. 

\begin{figure}[!htbp]
\begin{center}
    \includegraphics[scale=0.32]{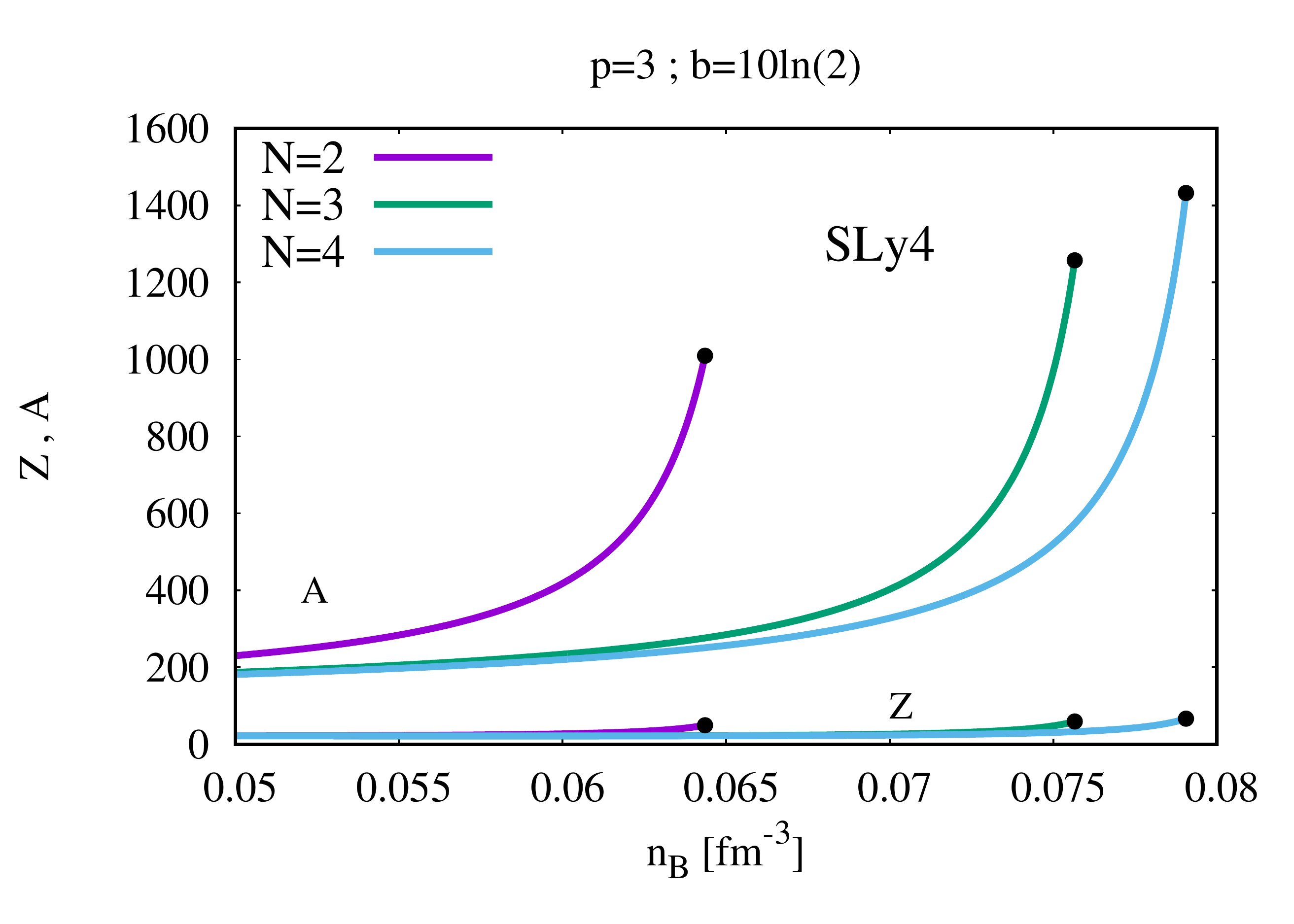}
    \includegraphics[scale=0.32]{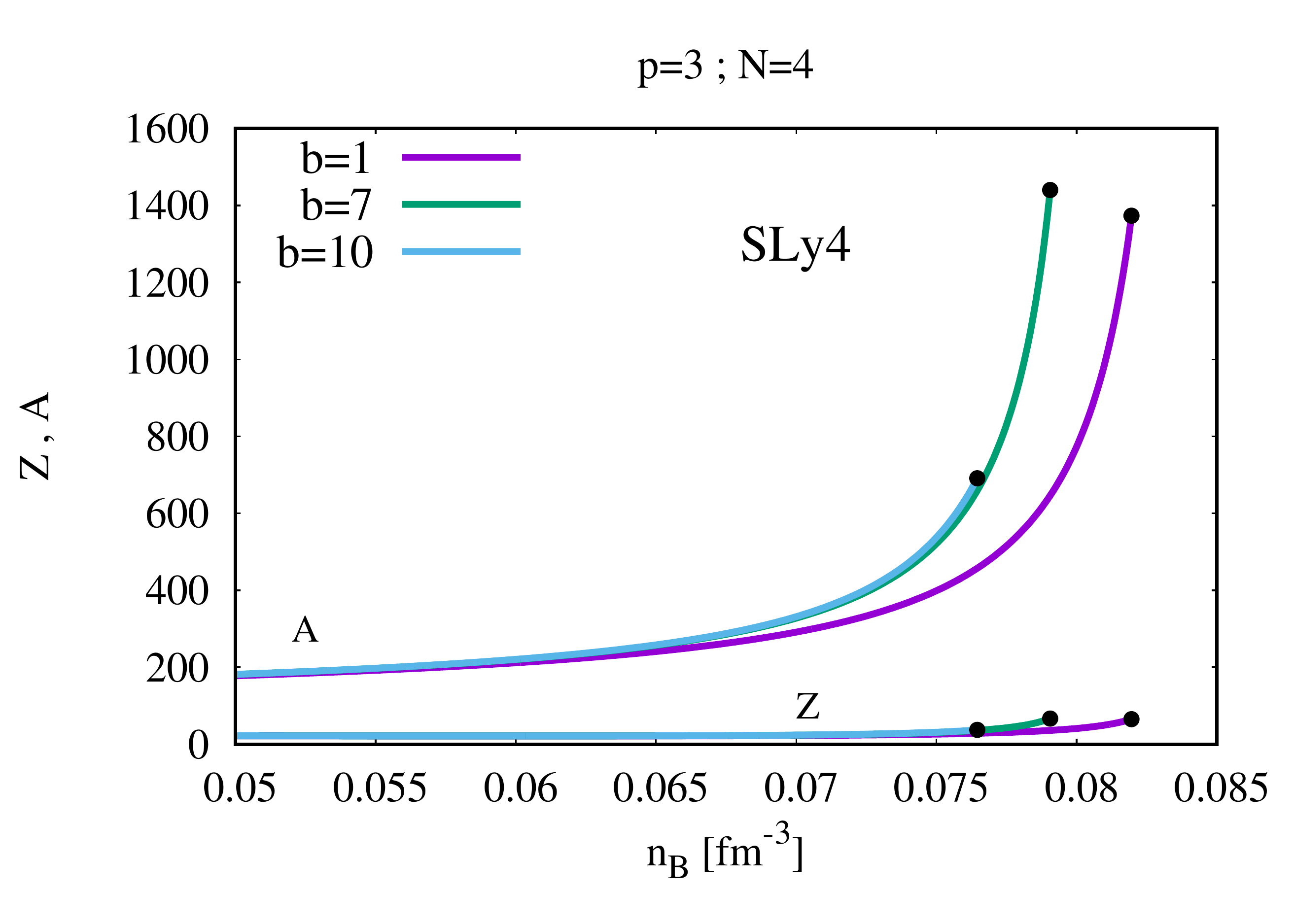}
    \includegraphics[scale=0.32]{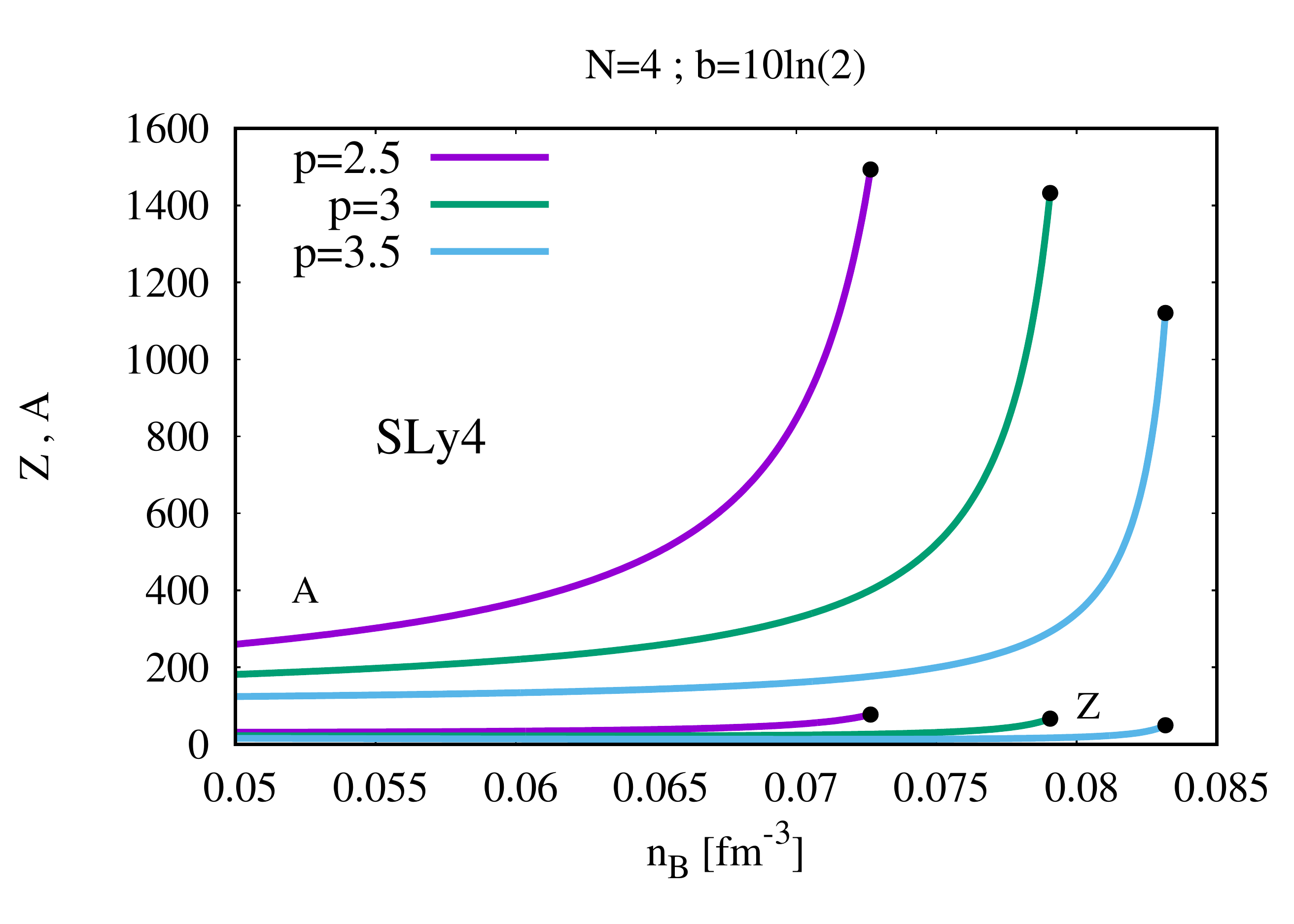}
\end{center}
    \caption{Crust composition for different orders of the density expansion (upper part), different values of the $b$ parameter (medium part), and different values of the surface tension $p$ parameter (lower part)  in the SLy4 meta-model. The ending point of the curves signals the transition point.}
\label{fig:nbp_influence}
\end{figure}

\subsection{Crust composition and CC transition point}

The influence of the uncontrolled parameters on the crust composition is analyzed in Fig.~\ref{fig:nbp_influence}. The upper part of the figure explores the importance of the high order terms in the density development. We can see that convergence is approximately reached at $N=3$, and a truncation at $N=2$ leads to an important underestimation of the transition density. This means that the $Q_q$ (and to a lesser degree $Z_q$) parameters cannot be neglected. $N=4$ will be used in the rest of the paper.

The medium panel in Fig.~\ref{fig:nbp_influence} explores the influence of the low density $b$ parameter. A very good reproduction of the SLy4 functional is obtained if $b=10\ln(2)$ is used. Taking different values of $b$ corresponds to considering EoS models which would have exactly the same empirical parameters (including effective masses and high order parameters up to $N=4$) as the SLy4 functional, but would differ in the treatment of the extreme low density domain. This is indeed what happens
with  the inclusion of deuteron or cluster correlations~\cite{Roepke2009,Roepke2015}. Since again the effect is non-negligible, we keep $b$ as an extra EoS parameter. The low density correction to the Taylor expansion around saturation induced by the $b$ term in eq.(\ref{eq:vELFc}) gets suppressed of a factor two at a density  $n_{min}/n_{sat}=\ln2/b$. We will consider in the following possible variations of $b/\ln 2$ in the interval $[1,10]$, which corresponds to a breaking down of the Taylor expansion at a density varying between $n_{min}=0.1 n_{sat}$ and $n_{min}= n_{sat}$.  This latter value corresponding to the minimum $b$ value might look quite extreme, but we will see in section~\ref{sec:bayes} that the influence of this parameter turns out to be negligible. 
Finally, the effect of varying the isovector surface $p$ parameter is reported in the lower panel of Fig.~\ref{fig:nbp_influence}.  Here the variation is done around the value $p=3$ used in the popular Lattimer and Swesty EoS~\cite{Lattimer1991}. We can see that both the average cluster size and the transition density are strongly influenced by the $p$ parameter, which cannot be directly linked to the uniform matter energy functional. From this observation we can already anticipate that the correlation between the CC transition and the EoS will  be considerably blurred by our lack of knowledge of the isovector surface tension. 

\begin{figure}[htbp]
    \centering
    \includegraphics[scale=0.39]{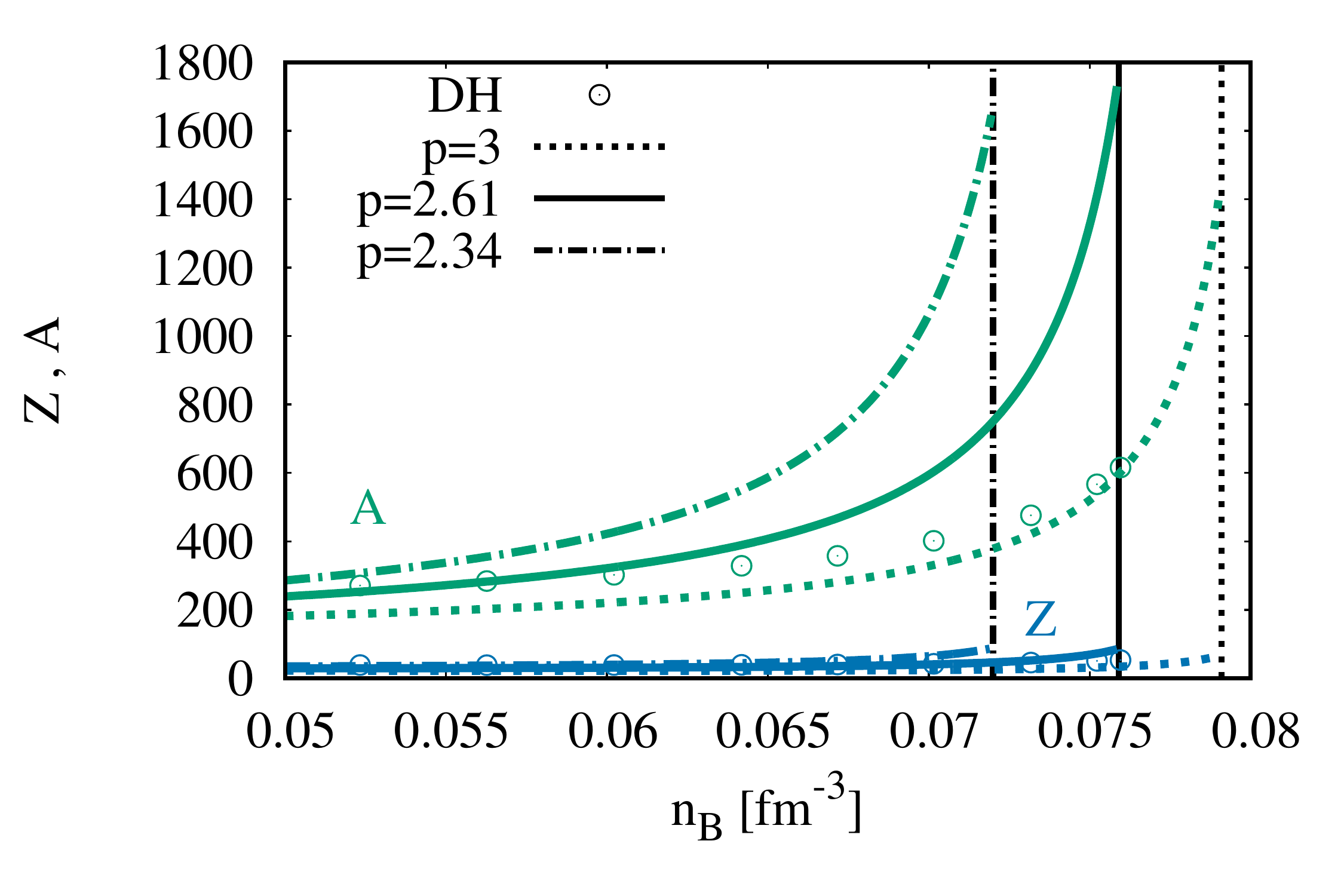}
    \caption{Inner-crust composition $Z$ (in blue) and $A$ (in green) as a function of the baryon density $n_B$. The dots are the results of the CLD model of Ref.~\cite{Douchin2001} using the SLy4 interaction. The dotted line represents the result given by our model for $p=3$. Composition for $p=2.61$ ($p=2.34$) to reproduce the transition density of Ref.~\cite{Douchin2001} (Ref.~\cite{Vinas2017}) is also represented.}
\label{fig:crust_DH}
\end{figure}

\subsection{Comparison with previous calculations}

Our results with the meta-functional optimized to give a good reproduction of the SLy4 model are compared with the other results available in the literature from a direct modelling of the inner crust using the SLy4 functional and different approximations for the surface tension~\cite{Douchin2001,Vinas2017}. In the case of the popular Douchin and Haensel (DH) model~\cite{Douchin2001}, we also report the average cluster size and charge for comparison with our results. We can see that the value of the cluster charge $Z$ is resonably compatible, but strong differences exist between the different calculations concerning the cluster size as well as the transition density. The difference in the cluster size between our approach and the DH one may be simply due to the different definitions of a cluster in a dense medium, which is affected by a certain degree of arbitrariness~\cite{Newton2013,Panagiota2013}. 
Conversely, the difference in the transition point is most probably due to the slightly different treatments of the isovector surface tension. We can see that choosing $p=2.61$ ($p=2.34$) allows reproducing the previous results by DH (Vinas et al.~\cite{Vinas2017}), and the difference between Ref.~\cite{Douchin2001} and Ref.~\cite{Vinas2017} is of the same order as the difference between our results using $p=2.61$ or the canonical value $p=3$ from Ref.~\cite{Lattimer1991}.

\begin{table}[]
    \centering
    \begin{tabular}{lccccc}
       \hline
        \hline
        \multicolumn{1}{c}{} & \multicolumn{5}{c}{$n_t$ (fm$^{-3}$)} \\
        Model &  $n_{td}$ &  $n_{tt}$ & $p=2.5$ & $p=3.0$ & $p=3.5$ \\
        \hline
        BSk14   & 0.081 & 0.090 & 0.073 & 0.079 & 0.085 \\
        BSk16   & 0.087 & 0.096 & 0.080 & 0.087 & 0.092 \\
        BSk17   & 0.086 & 0.095 & 0.078 & 0.085 & 0.091 \\
        NRAPR   & 0.073 & 0.083 & 0.062 & 0.071 & 0.076 \\
        RATP    & 0.086 & 0.097 & 0.078 & 0.087 & 0.092 \\
        SkO     & 0.062 & 0.073 & 0.050 & 0.061 & 0.064 \\
        SLy230a & 0.081 & 0.090 & 0.076 & 0.080 & 0.085 \\
        SLy230b & 0.080 & 0.089 & 0.073 & 0.079 & 0.083 \\
        SLy4    & 0.080 & 0.089 & 0.073 & 0.079 & 0.083 \\
        NL3     & 0.054 & 0.065 & 0.046 & 0.054 & 0.061 \\
        TM1     & 0.060 & 0.070 & 0.049 & 0.058 & 0.062 \\
        DD-ME1  & 0.070 & 0.085 & 0.064 & 0.076 & 0.083 \\
        DD-ME2  & 0.072 & 0.087 & 0.071 & 0.081 & 0.087 \\  
        \hline
        \hline
    \end{tabular}

    \vspace{0.4cm}
    \begin{tabular}{lccccc}
        \hline
        \hline
        \multicolumn{1}{c}{} & \multicolumn{5}{c}{$P_t$ (MeV/fm$^3$)} \\
        Model &  $P_{td}$ &  $P_{tt}$ & $p=2.5$ & $p=3.0$ & $p=3.5$ \\
        \hline
        BSk14   & 0.381 & 0.483 & 0.311 & 0.366 & 0.433 \\
        BSk16   & 0.402 & 0.502 & 0.340 & 0.409 & 0.459 \\
        BSk17   & 0.397 & 0.499 & 0.324 & 0.391 & 0.455 \\
        NRAPR   & 0.413 & 0.545 & 0.299 & 0.391 & 0.454 \\
        RATP    & 0.390 & 0.500 & 0.321 & 0.401 & 0.452 \\
        SkO     & 0.270 & 0.413 & 0.162 & 0.271 & 0.315 \\
        SLy230a & 0.319 & 0.404 & 0.269 & 0.307 & 0.351 \\
        SLy230b & 0.362 & 0.462 & 0.296 & 0.355 & 0.397 \\
        SLy4    & 0.361 & 0.461 & 0.296 & 0.355 & 0.397 \\
        NL3     & 0.236 & 0.422 & 0.160 & 0.261 & 0.368 \\
        TM1     & 0.324 & 0.511 & 0.177 & 0.302 & 0.362 \\
        DD-ME1  & 0.404 & 0.605 & 0.391 & 0.526 & 0.607 \\
        DD-ME2  & 0.409 & 0.594 & 0.445 & 0.550 & 0.616 \\  
        \hline
        \hline
    \end{tabular}
    \caption{Transition density $n_t$ (top) and transition pressure $P_t$ (bottom) for several interactions. 
Meta-modelling unified EoS calculations with $p=2.5$, $p=3$, $p=3.5$ are given together with the compilation by Ducoin \textit{et al.} from Ref.~\cite{Ducoin2011} using the dynamical ($n_{td},P_{td}$) and thermodynamical ($n_{tt},P_{tt}$) methods.}
\label{tab:ducoin}
\end{table}


As we have already discussed in the introduction, a very limited number of works exists computing the CC transition from a direct modelling of the inner crust, because of the complexity of the simulations. A greater effort has been devoted to the calculations from the core side, using the thermodynamical or dynamical spinodal technique. The compilation of Ref.~\cite{Ducoin2011} of the transition density and pressure calculated for different models is reported in Table~\ref{tab:ducoin} 
. For each model, the results compiled in Ref.~\cite{Ducoin2011} are compared to our calculation where the parameter set $\vec X$ is fixed such as to reproduce the considered model. A part of these results can also be found in Ref.~\cite{Thomas_prl}. 

We can see that the transition density results of the literature for the dynamical spinodal are globally nicely reproduced by our calculation "from the crust" with the choice $p=3$. Higher values of the isovector surface tension parameter are needed if we want to reproduce the estimations of the thermodynamical spinodal. 
The transition pressure is more fluctuating, but still a variation of $p$ in the interval $[2.5,3.5]$ allows reproducing all the considered models. A flat probability distribution within this interval will be our prior for the statistical analysis of section~\ref{sec:bayes}.

 \begin{figure}[htbp]
    \centering
    \includegraphics[scale=0.29]{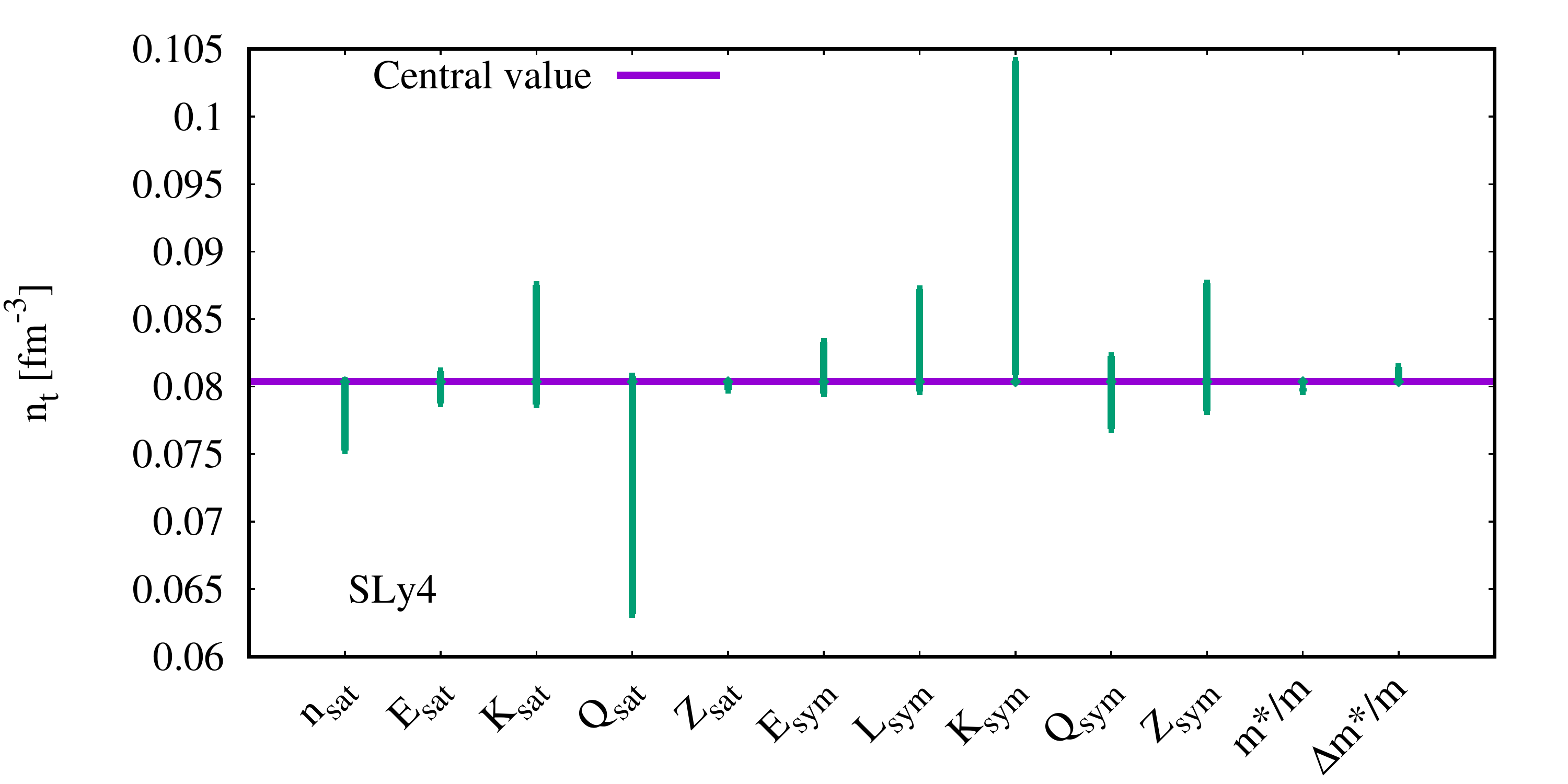}
    \includegraphics[scale=0.29]{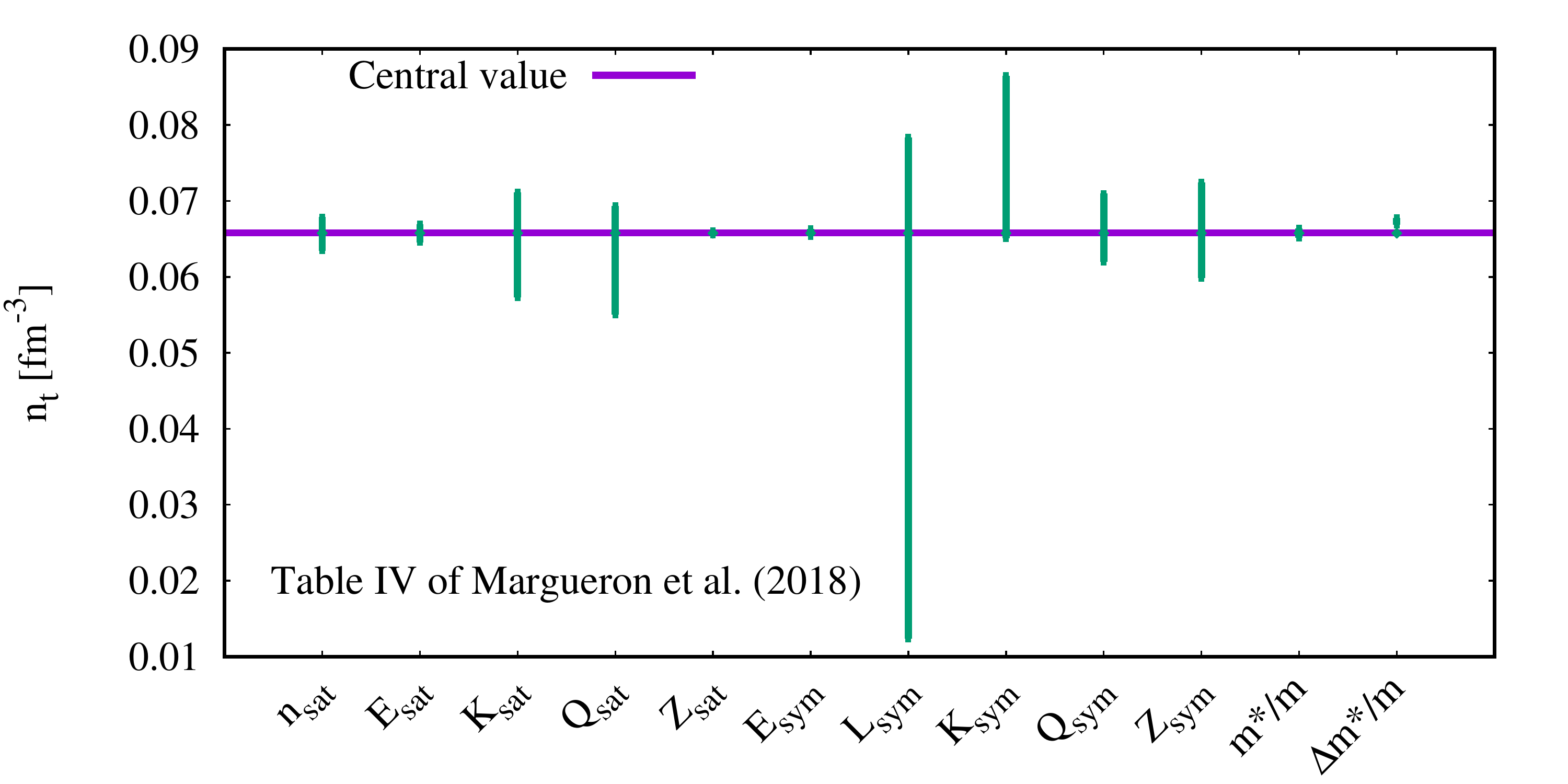}
    \caption{Sensitivity analysis of the transition density $n_t$ with respect to EoS parameters. $p=3$ is fixed and two different reference points are chosen: SLy4 parameters (top) and parameters of Table IV of Ref.~\cite{Margueron2018a} (bottom).}
\label{fig:sensana_nt}
\end{figure}

\section{Sensitivity analysis}\label{sec:sensitivity}

\begin{figure}[htbp]
    \centering
    \includegraphics[scale=0.29]{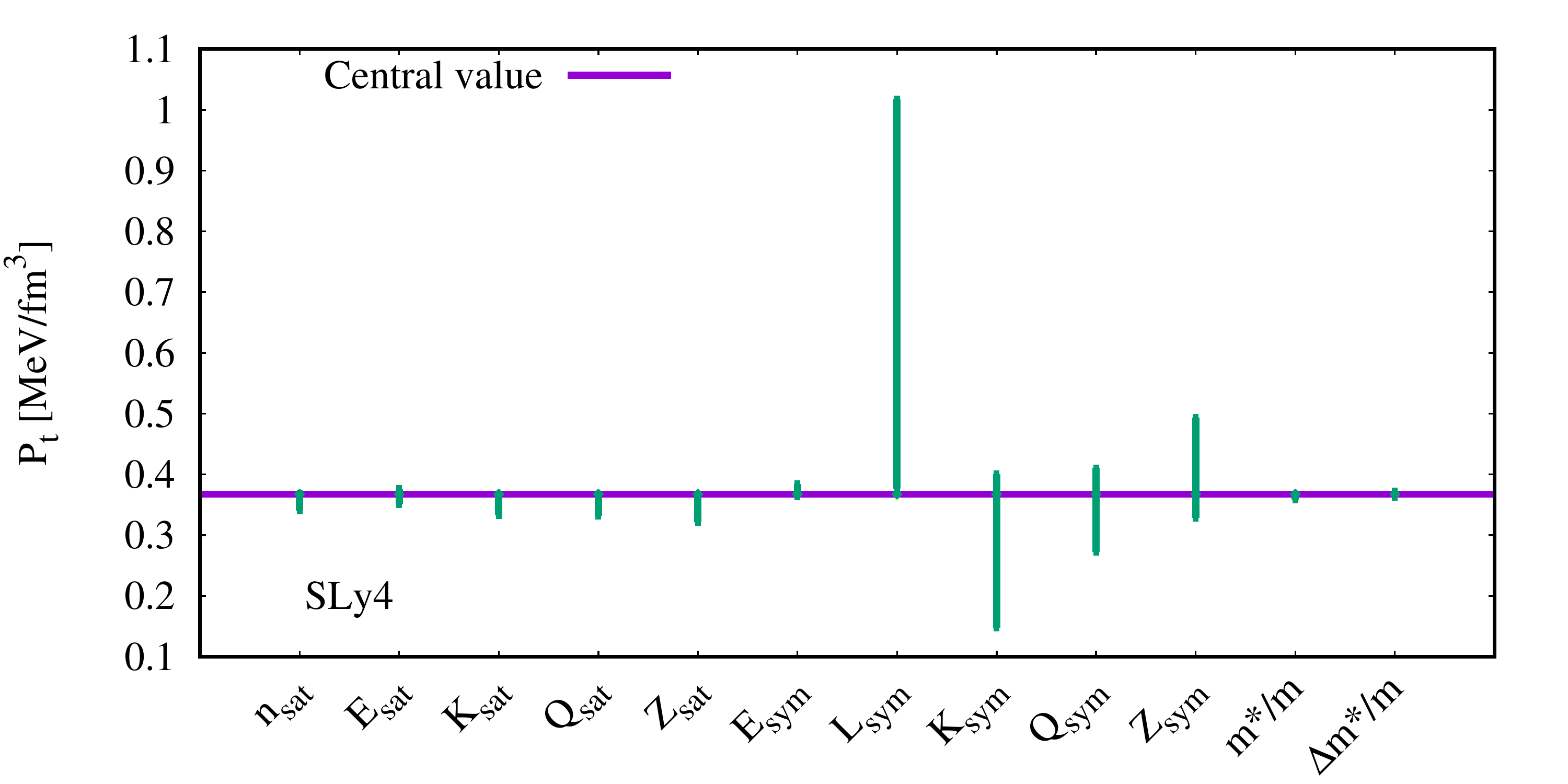}
    \includegraphics[scale=0.29]{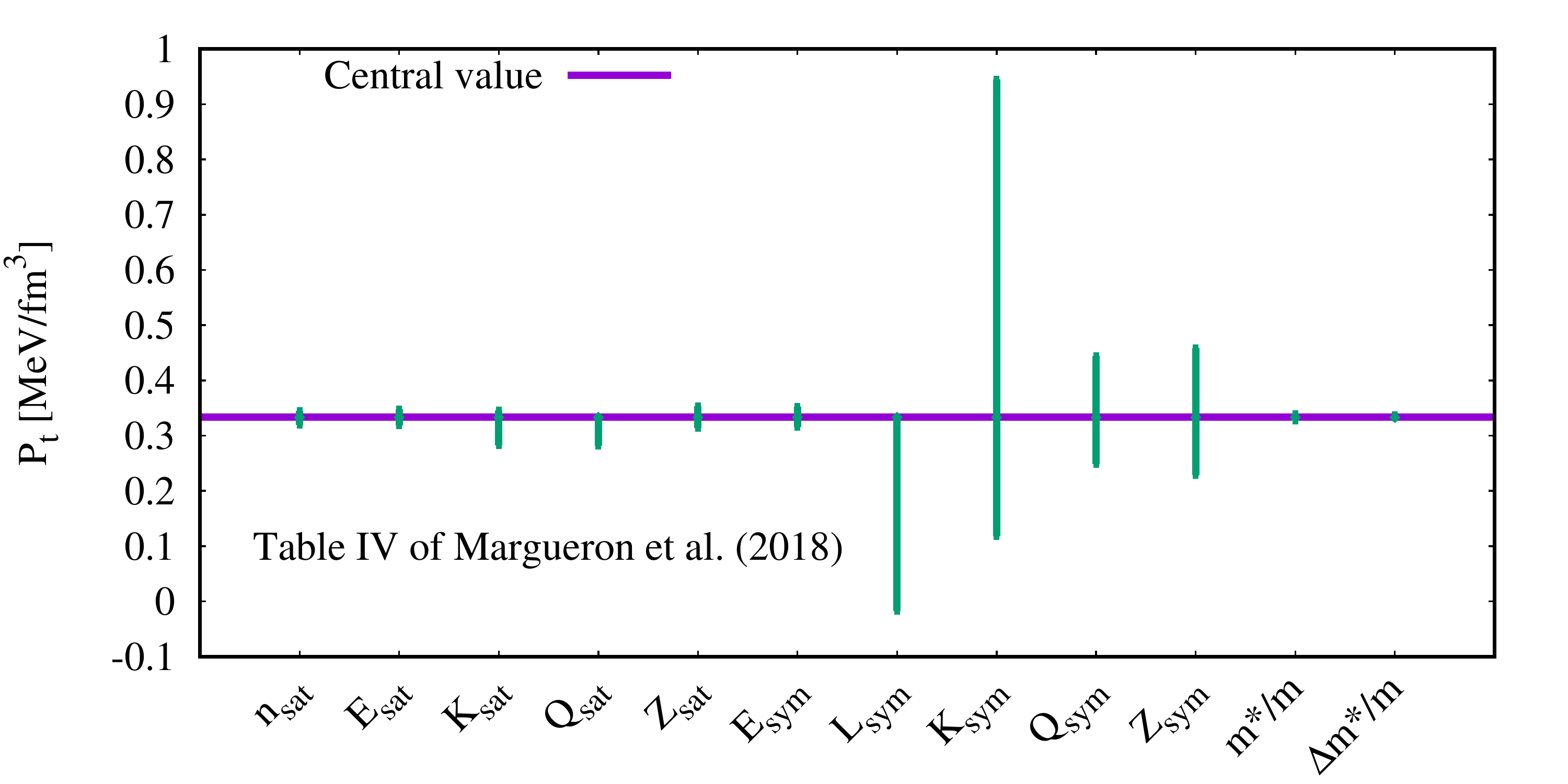}
    \caption{Same as Fig.~\ref{fig:sensana_nt} for the transition pressure $P_t$.}
\label{fig:sensana_pt}
\end{figure}

One of the advantages of the meta-modelling technique is that, since all the model parameters are a-priori uncorrelated, it is possible to vary each one of them independently of the others, which is not possible using specific functional behaviors such as Skyrme or Gogny or the different versions of RMF. This allows determining the most influential parameters on any given observable. Such a sensitivity analysis is presented in Fig.~\ref{fig:sensana_nt} for the transition density and in
Fig.~\ref{fig:sensana_pt} for the transition pressure. 

The one-by-one variation of all the EoS parameters is performed around two different reference parameter set $\vec X_{ref}$, namely the parameter set corresponding to the SLy4 model (upper part of Figs.~\ref{fig:sensana_nt},\ref{fig:sensana_pt}), and the set of average values of the different parameters from the compilation of empirical constraints in Ref.~\cite{Margueron2018a} (lower part of Figs.~\ref{fig:sensana_nt},\ref{fig:sensana_pt}).

\begin{table}[]
\centering
\label{my-label}
\begin{tabular}{cccccccc}
\hline\hline
\multicolumn{1}{c|}{\multirow{2}{*}{Parameter}} & \multicolumn{1}{c|}{\multirow{2}{*}{Unit}} & \multicolumn{2}{c|}{Prior}                         & \multicolumn{2}{c|}{HD}                         & \multicolumn{2}{c}{LD}                         \\ \cline{3-8} 
\multicolumn{1}{c|}{}                  & \multicolumn{1}{c|}{}                  & \multicolumn{1}{c}{Min} & \multicolumn{1}{c|}{Max} & \multicolumn{1}{c}{Average} & \multicolumn{1}{c|}{$\sigma$} & \multicolumn{1}{c}{Average} & \multicolumn{1}{c}{$\sigma$} \\ \hline
$n_{sat}$ & fm$^{-3}$ & 0.15 & 0.17 & 0.1600 & 0.0060  & 0.1641  & 0.0049 \\
$E_{sat}$ & MeV & -17 & -15  & -16.01  & 0.61  & -15.29  & 0.25 \\
$K_{sat}$ & MeV & 190 & 270 & 229  & 24  & 234  & 23 \\
$Q_{sat}$ & MeV & -1000 & 1000 & 200  & 535  & -31  & 362 \\
$Z_{sat}$ & MeV & -3000 & 3000 & 1038  & 1233 & -146  & 1728 \\
$E_{sym}$ & MeV & 26  & 38  &  33.53 & 3.48  & 30.71  & 0.76 \\
$L_{sym}$ & MeV & 10  &  80 & 45.45  & 17.97  & 43.66  & 3.68 \\
$K_{sym}$ & MeV & -400  & 200  & -92  &  136 & -202  & 42 \\
$Q_{sym}$ & MeV & -2000  & 2000  & 913  & 740  & -253  & 673 \\
$Z_{sym}$ & MeV & -5000  & 5000  & 1463  & 2216  & -114  & 2868 \\
$m^*_{sat}/m$ & & 0.6  & 0.8 & 0.70  & 0.06  & 0.70  & 0.06 \\
$\Delta m^*_{sat}/m$ & & 0.0  & 0.2 & 0.10  & 0.06  & 0.10  & 0.06 \\
$b$ &   & 1  & 10  & 5.3  & 2.7  & 5.2  & 2.6 \\ \hline\hline
\end{tabular}
\caption{Minimum value and maximum value of each of the empirical parameters for the prior distribution (prior) and average and standard deviation of each of the empirical parameters of the posterior distribution after application of the HD(LD) filter (see text).} \label{tab:param}
\end{table}

The minimum and maximum value chosen for each parameter are taken from Ref.~\cite{Margueron2018b} and they are given in Table~\ref{tab:param}. These values reflect the degree of uncertainty on the different parameters, as measured by their observed variation in the different functionals that have been successfully confronted to low energy nuclear physics data. The vertical lines in Figs.~\ref{fig:sensana_nt},\ref{fig:sensana_pt}) give the transition density and pressure domain obtained when the EoS parameters are one by one varied around the reference model, within the interval of Table~\ref{tab:param}. Since the uncertainty on the different parameters is not the same, the length of the segments is a qualitative measure of the propagation of the uncertainty on the transition point brought by each parameter.   

We can see that the sensitivity of each parameter depends on the value of the other parameters, that is on the chosen reference set $\vec X_{ref}$. Still, universal trends clearly emerge. We can see that the CC phase transition, at variance with the standard liquid-gas of symmetric matter, is virtually insensitive to isoscalar parameters. Even if extremely large variations of $Q_{sat}$ and $Z_{sat}$ are considered (see Table~\ref{tab:param}), the prediction of the transition point is almost unaffected. This underlines the importance of the energetics of the neutron gas on the transition point. Concerning the isovector sector, we can see that $L_{sym}$ is the most important parameter. This result is in agreement with previous findings by many authors~\cite{Ducoin2011}. The symmetry energy at saturation $E_{sym}$ and the effective mass splitting $\kappa_v$ do not play any role on the transition, which can be partially explained by the fact that these parameters are already relatively well constrained. Depending on the chosen reference point, the transition pressure shows also a great sensitivity to the isovector compressibility $K_{sym}$. This can explain why the transition pressure exhibits an irregular behavior when plotted as a function of $L_{sym}$~\cite{Ducoin2011}: the different functionals considered in the litterature have very different values of $K_{sym}$, which blurs the correlation with $L_{sym}$. This effect is also amplified by the fact that, depending on the reference point, the dependence of $P_t$ with $L_{sym}$ is not monotonic. Finally, we can remark that the influence of the fully unknown high order derivatives $Q_{sym}$ and $Z_{sym}$, though less important than the one of $L_{sym}$, is not negligible and comparable to the one of the isovector surface energy parameter $p$ that can be inferred from Table~\ref{tab:ducoin}.
Similar conclusions can be drawn if the sensitivity analysis is performed using the definition of the transition point from the dynamical spinodal~\cite{Chatterjee,Antic}. 

\section{Statistical analysis}\label{sec:bayes}

\begin{figure}[htbp]
\begin{center}
    \includegraphics[scale=0.68]{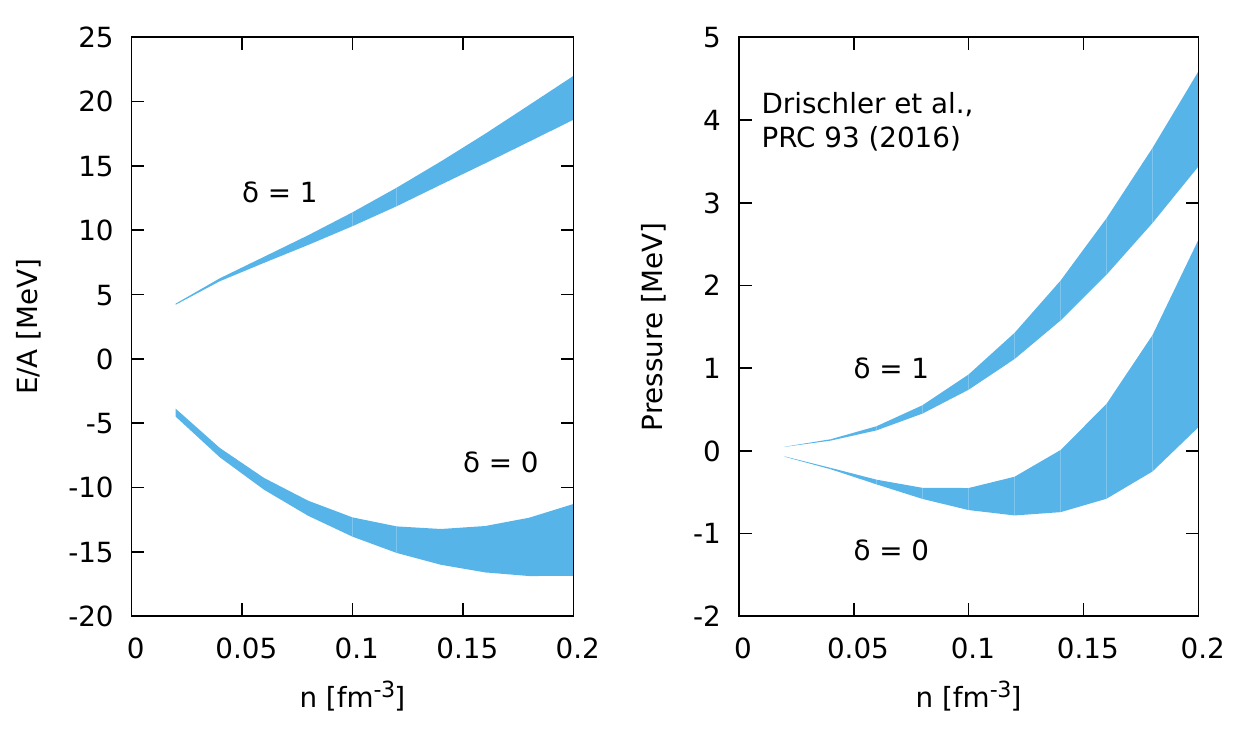}
\end{center}
\caption{Low density constraints (energy per nucleon and pressure) for neutron and symmetric matter from the EFT ab-initio calculation of Ref.~\cite{Drischler2016}. Blue bands are associated to our LD filter.}
\label{fig:EFT1}
\end{figure}

We now come to the quantitative determination of the CC transition point and its uncertainty, on the basis of our imperfect knowledge of the nuclear energy functional. To this aim, we perform a Bayesian determination of the model parameters on the full $2(N+1)+3$ parameter space, up to the fourth order in the Taylor expansion ($N=4$).
The prior distribution of $\vec X$ is given by an uncorrelated ansatz and a flat distribution of each parameter within the interval specified in Table~\ref{tab:param},
\begin{equation}
p_{prior}(\vec X) = \prod_{k=1}^{2(N+1)+3}f(X_k^{min},X_k^{max};X_k) \label{eq:prior}
\end{equation}
where $f$ is a uniform distribution between $X_k^{min}$ and $X_k^{max}$, defined in terms of the Heavyside step function $H(x)$ as $f(a,b;x)=(H(x-a)-H(x-b))/(b-a)$.
The posterior distribution is obtained by applying different physical filters to the prior distribution,
\begin{eqnarray}
p_{post} (\vec X) = \mathpzc{N} \, w_{LD(HD)}(\vec X) \, e^{-\chi^2(\vec X)/2} \, p_{prior}(\vec X)  .
\label{eq:probalikely}
\end{eqnarray}
In this expression, both strict ($w$ term) and likelyhood (exponential term) filters are applied, and  $\mathpzc{N}$ is a normalization. $\chi^2(\vec X)$ represents the $\chi^2$ corresponding to the optimal fit of nuclear masses which is done to determine the surface tension parameters $\sigma_0$ and $b_s$ for each $\vec X$ parameter set (see section~\ref{sec:surface}). $w$ is a sharp $\delta$-function filter that outputs 1 if the constraint is respected, and 0 otherwise. 

Two different constraints are considered for the $w$ filter. 
The first constraint, noted $w_{LD}$, concentrates on the low density (LD) $n\leq n_{sat}$ behavior of the energy functional. We impose to the different functionals generated following eq.(\ref{eq:prior}) to strictly pass through the uncertainty band of the N3LO effective field theory calculation for symmetric and pure neutron matter by  Drischler et al.~\cite{Drischler2016}. This same condition was applied in previous studies~\cite{Thomas_prl,Chatterjee,Antic,Holt}.
This filter is applied in the density interval $[0.05,0.2]$ fm$^{-3}$. The very low density region is not considered because
of numerical issues due to the very small uncertainty, but we recall that all the generated models by construction converge to zero energy and pressure in the $n\to 0$ limit. The energy per particle and pressure uncertainty bands applied are displayed in Fig.~\ref{fig:EFT1}. In this figure, the interval delimited by the dotted lines corresponds to our prior distribution. We can see that the uncertainty band of the prior energy per particle and pressure of symmetric matter around saturation is
comparable and even narrower than the one corresponding to the ab-initio calculation. This is due to the strong empirical constraints coming from different low energy nuclear physics experiments, that have been considered to determine realistic intervals for the empirical parameters in Table~\ref{tab:param}. Conversely, as it is well known, the empirical information embedded in the prior distribution is insufficient to effectively constrain the neutron matter EoS, and the ab-initio predictions
are much narrower than our prior distribution. In the isovector sector the LD filter is extremely selective: only 2118 models out of the 100 millions generated to numerically sample the prior parameter distribution,  fulfill the LD condition. 

The second (HD) filter imposes general physical constraints to the global density behavior of the functional, as follows:
 
\begin{itemize}
    \item positive symmetry energy at all densities,
    \item $P(n)\geq 0$ for $n\geq n_{sat}$ (stability of the EoS),
    \item $0 < v_s < c$,
    \item $M_{max} > 2M_\odot$,
\end{itemize}

where $M_{max}$ is the maximum neutron star mass obtained through a TOV calculation~\cite{Haensel_book}.

Once the posterior parameter distribution is determined, the probability distribution of any parameter or observable $Y$ can be straightforwardly computed as:

\begin{equation}
p(Y) = \prod_{k=1}^{2(N+1)+3} \int_{X_k^{min}}^{X_k^{max}} dX_k Y(\vec X) p_{post}(\vec X),
\label{eq:distri}
\end{equation}

where $ Y(\vec X)$ is the value of the Y variable as obtained with the $\vec X$ parameter set.

\subsection{Distribution of  parameters and observables}

\begin{figure}[htbp]
    \centering
    \includegraphics[scale=0.44]{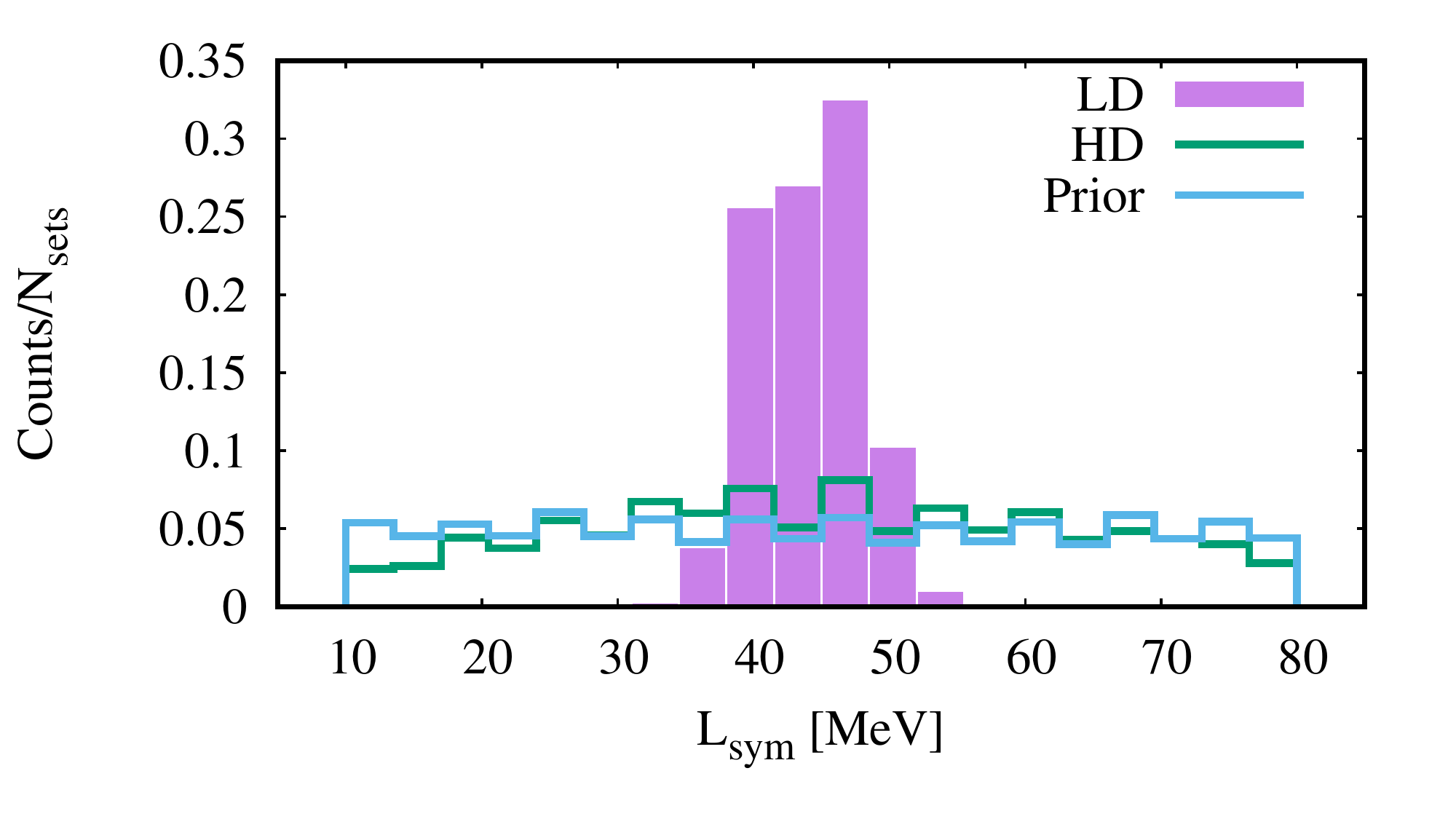}
    \includegraphics[scale=0.44]{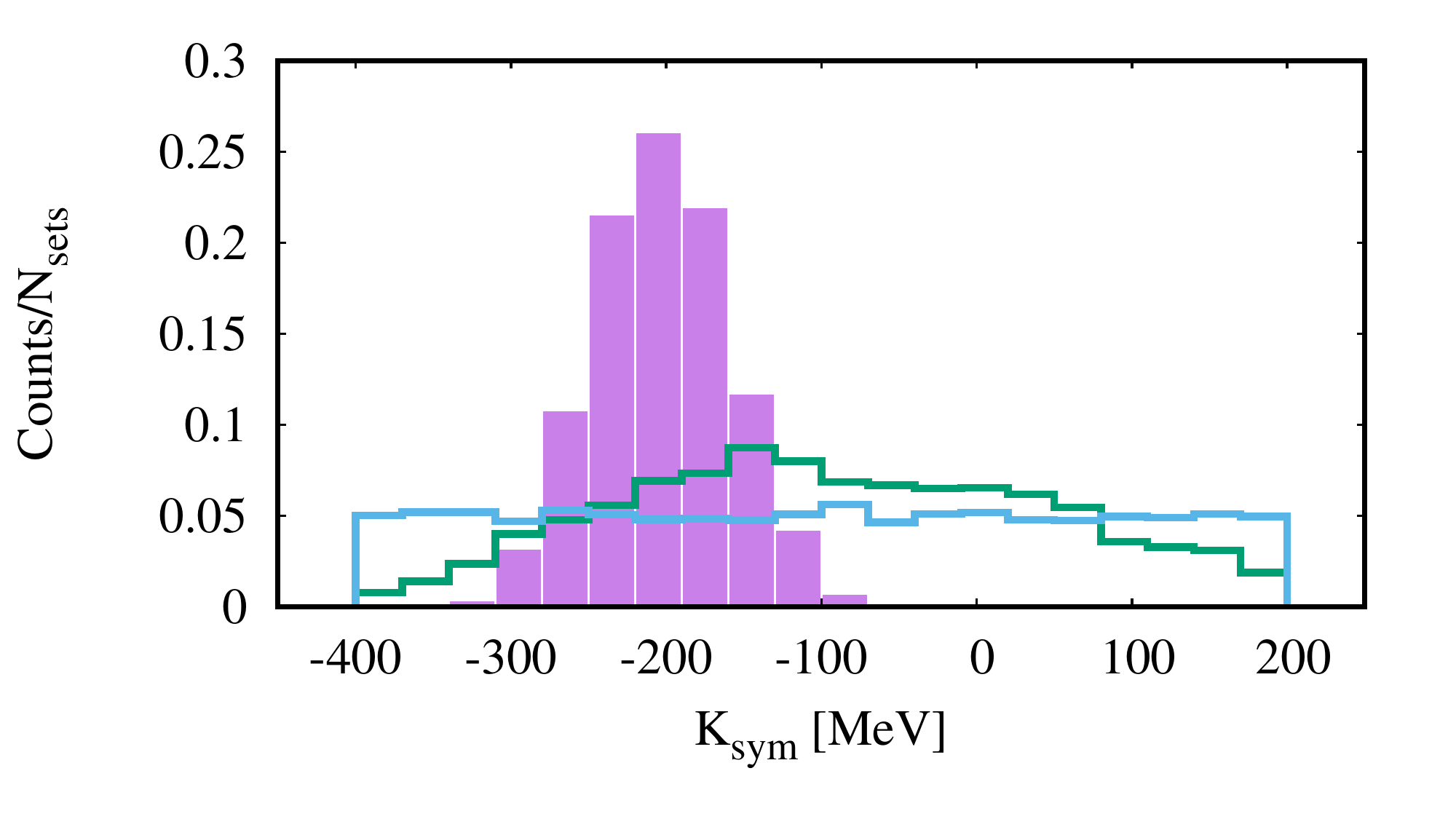}
    \includegraphics[scale=0.44]{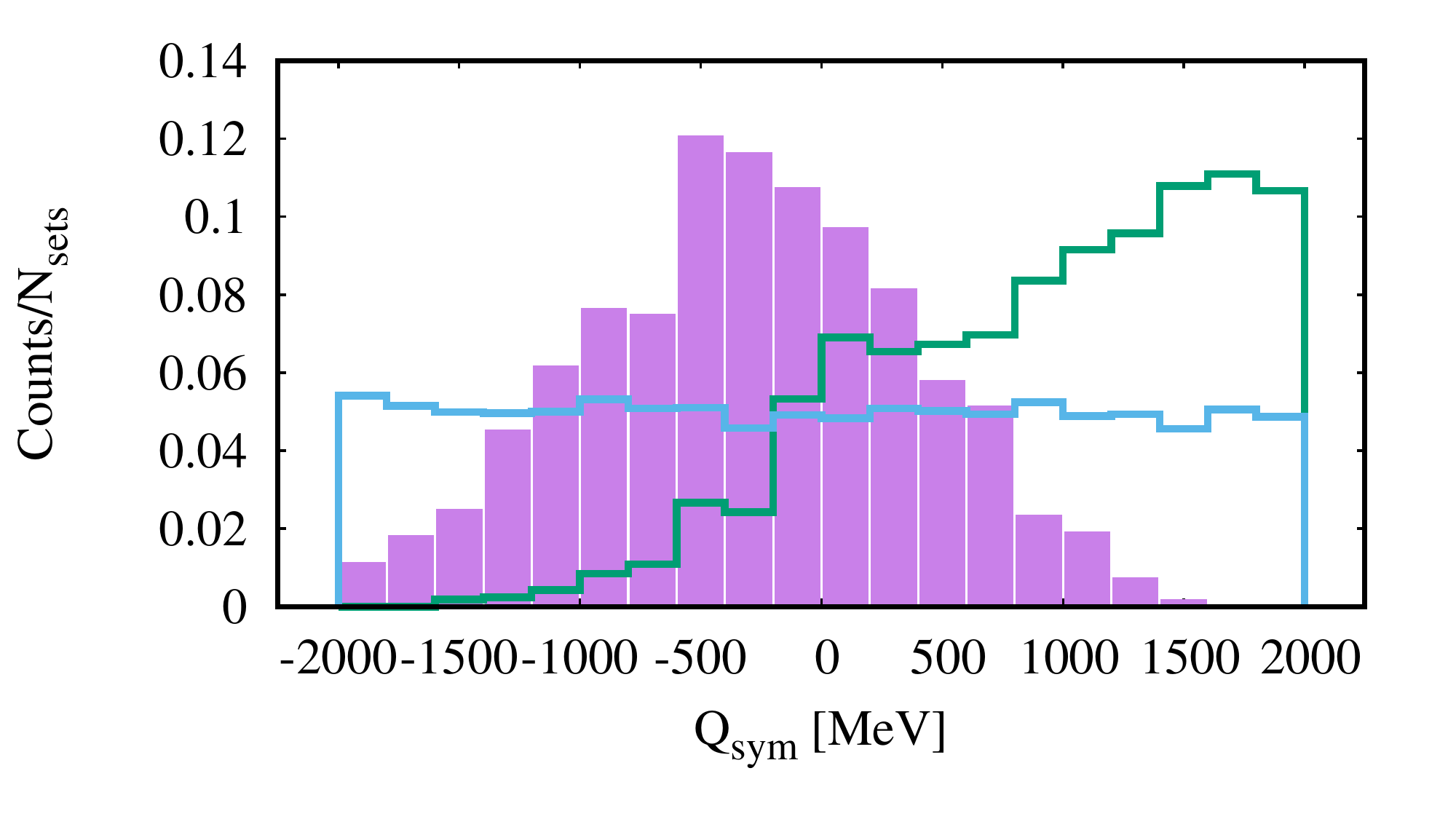}
    \caption{Distributions of $L_{sym}$ (top), $K_{sym}$ (middle), and $Q_{sym}$ (bottom) for the sets passing through the LD constraints (purple), HD constraints (green), and for random sets (blue).} \label{fig:param_post}
\end{figure}

The posterior distribution of the most influential isovector EoS parameters is displayed in Figure~\ref{fig:param_post}.
We can see that the general physical conditions corresponding to the HD filter almost do not constrain the low order empirical parameters, in agreement with the findings of Ref.~\cite{Margueron2018b}. Conversely, the LD filter allows a very tight determination of the empirical parameters $L_{sym}$ and $K_{sym}$.

\begin{figure}[htbp]
    \centering
    \includegraphics[scale=0.44]{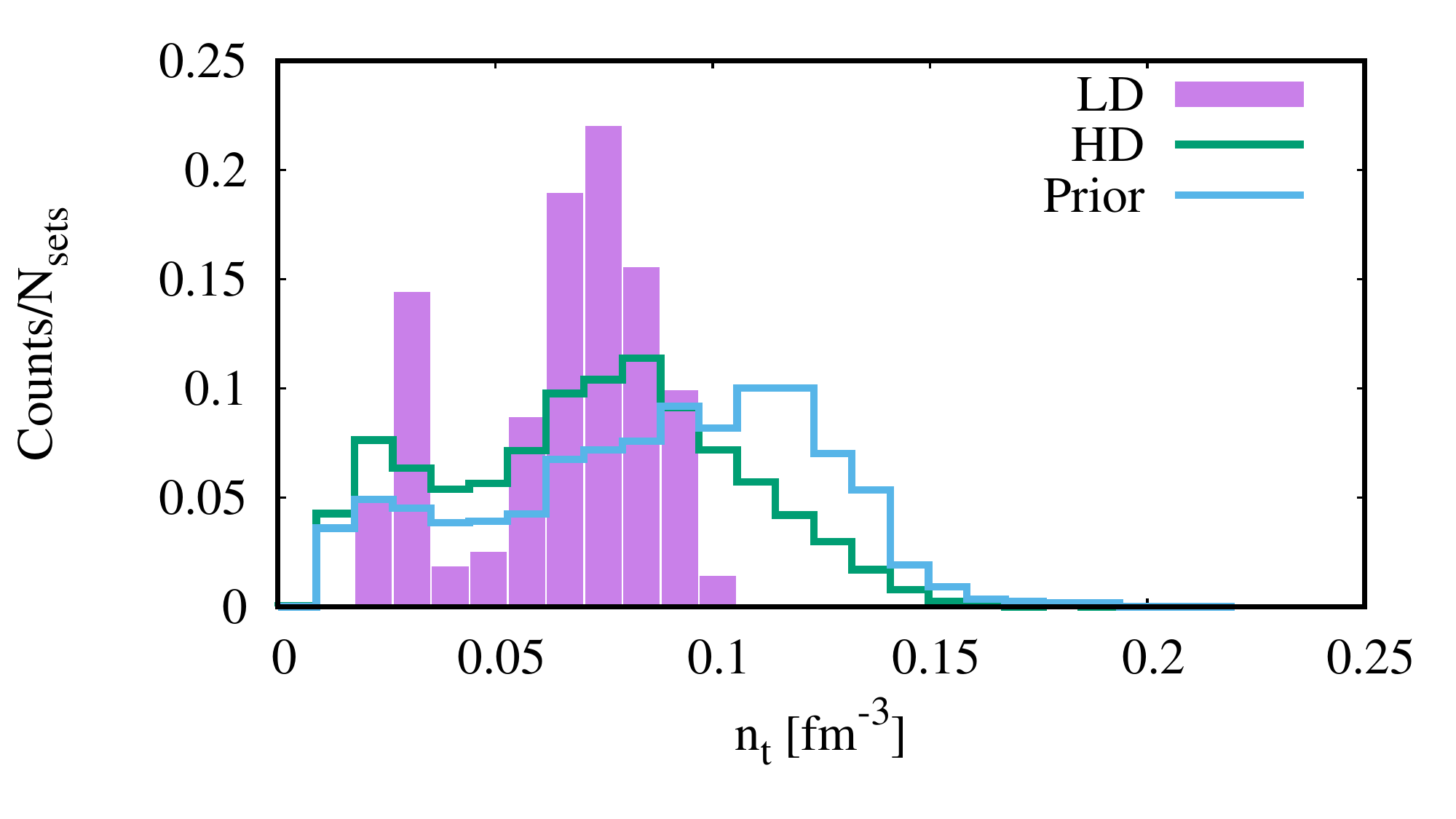}
    \includegraphics[scale=0.44]{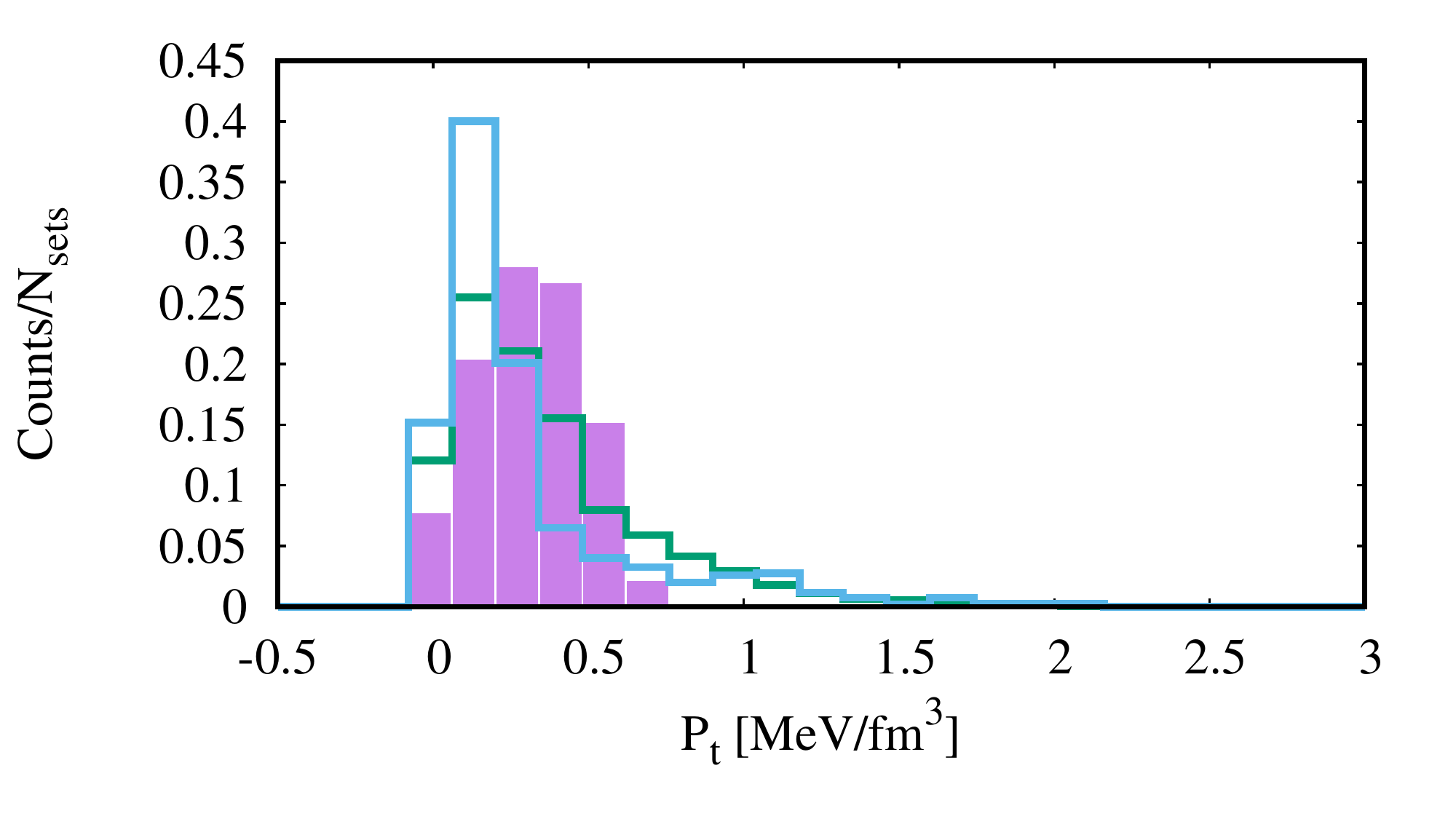}
    \caption{Distributions of $n_t$ (top), and $P_t$ (bottom) for the sets passing through the LD constraints (purple), HD constraints (green), and for random sets (blue).} \label{fig:ntpt_post}
\end{figure}

The effect on the distribution of the transition density and pressure is displayed in Fig.~\ref{fig:ntpt_post}.
The two-humped posterior distribution is due to the isovector surface energy parameter $p$, for which we consider for simplicity only three values 2.5, 3, and 3.5 with equal probability: 
 the left peak in the distribution of $n_t(P_t)$ is associated to $p=2.5$ while we can not distinguish between $p=3$ and $p=3.5$. This shows that the highest uncertainty in the determination of the transition point, once the EoS is constrained through the most advanced ab-initio calculations of nuclear matter, is due to our poor knowledge of the surface properies of extremely neutron rich matter.

\begin{table}[]
\centering
\label{my-label}
\begin{tabular}{lcccc}
\cline{2-5}
\hline\hline
\multicolumn{1}{l}{} & \multicolumn{2}{c|}{$n_t$ (fm$^{-3}$)}                         & \multicolumn{2}{c}{$P_t$ (MeV/fm$^3$)}                         \\ 
\multicolumn{1}{c}{} & \multicolumn{1}{c}{Average} & \multicolumn{1}{c|}{$\sigma$} & \multicolumn{1}{c}{Average} & \multicolumn{1}{c}{$\sigma$} \\ \cline{2-5} 
\hline
Prior ($p=3$) & 0.078 & 0.040 & 0.342 & 0.426 \\
HD ($p=3$) & 0.076 & 0.032 & 0.394 & 0.327 \\
LD ($p=3$) & 0.074 & 0.011 & 0.360 & 0.122 \\ 
LD & 0.065 & 0.021 & 0.307 & 0.167 \\ 
\hline\hline
\end{tabular}
\caption{Average value and standard deviation of the transition density (pressure) with the prior parameter distribution, the posterior using the HD filter, and the posterior using the LD filter for $p=3$ and $p=\{2.5, 3, 3.5\}$.} \label{tab:ntpt}
\end{table}

This statement can be quantified by calculating the first  moments of the distribution.
We give in Table~\ref{tab:ntpt} the average value and the standard deviation of the transition density (pressure) $n_t(P_t)$ for the prior and posterior distributions. In each case, the convergence of the results with the number of sampled models fulfilling the most restrictive LD filter, is also checked. We can see that, if the value of the isovector surface tension parameter $p$ is fixed, the LD filter is much more effective in reducing the uncertainty on the transition point, with respect to the HD filter. This underlines the importance of precise constraints on the low density EoS parameters for a reliable prediction of neutron star crust properties, as it has been often stressed in the literature. If the transition point is computed with reasonably well-behaved models for neutron stars fulfilling the HD conditions and with empirical parameters within the accepted bands from nuclear physics experiments (see Table~\ref{tab:param}), without the more precise constraints from ab-initio calculations, the transition pressure can be badly predicted even in average. 

This discussion ignores the uncertainty that we have on the behavior of the surface tension for extreme isospin values. If we incorporate that uncertainty considering $p$ as an extra parameter (last line in Table~\ref{tab:ntpt}), we can see that, according to the range assumed for the prior distribution of $p$, the uncertainty on the transition point is considerably increased and even the average value is affected.

\subsection{Correlations among the empirical parameters and the transition}




\begin{figure}[htbp]
    \centering
    \includegraphics[scale=0.25]{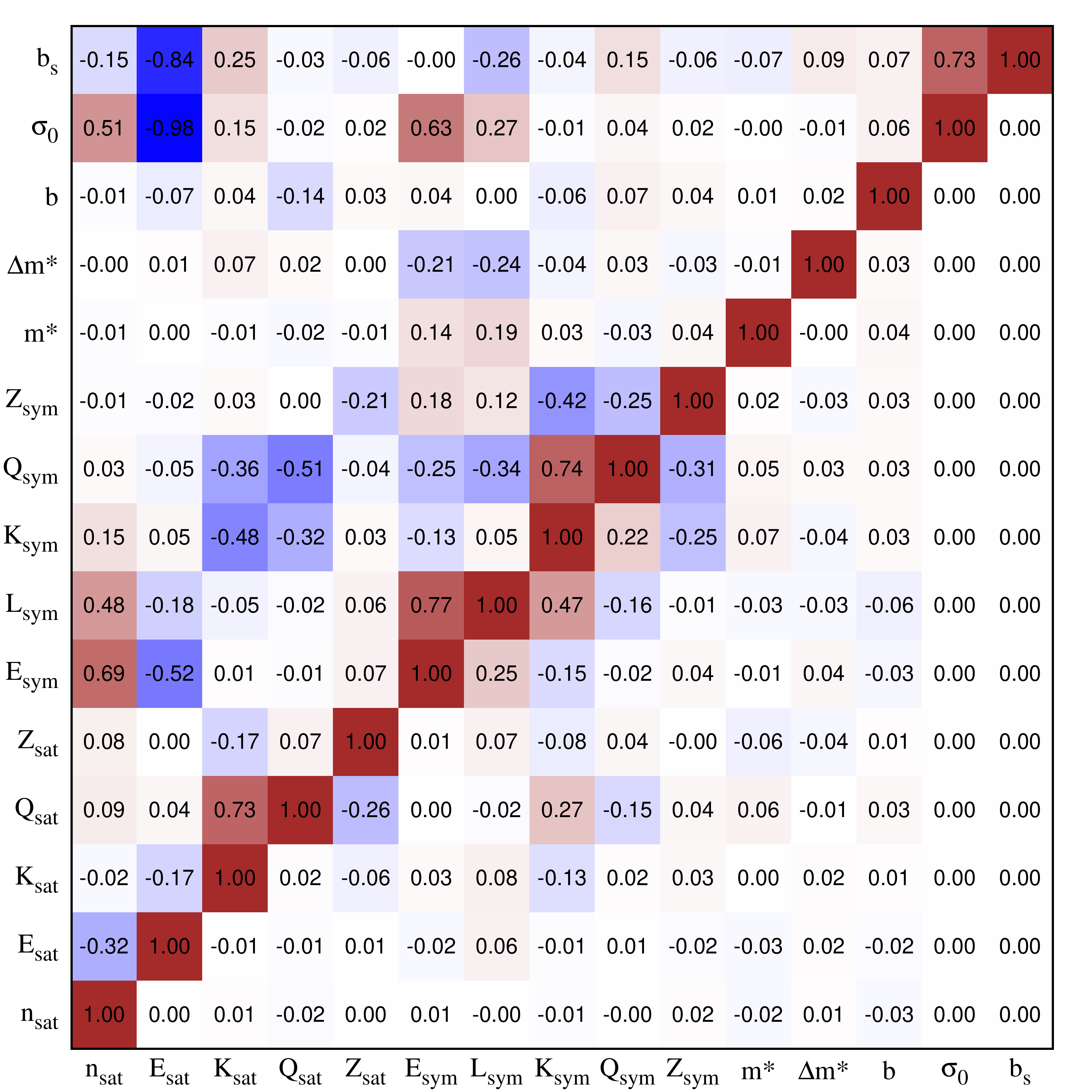}
    \caption{Correlation matrix for the empirical parameters and the surface parameters $\sigma_0$, $b_s$. The part under the diagonal show to the correlation coefficients for the sets passing through the HD filter only while the part above the diagonal corresponds to the correlation coefficients for the sets passing through the LD + mass filter.
}\label{fig:cormat}
\end{figure}

We now turn to explore the correlations between the empirical EoS parameters and the transition point, as obtained by applying the different filters.
Fig.~\ref{fig:cormat} displays the correlation matrix among the different EoS parameters. The isovector surface tension parameter $p$ is fixed to $p=3$ for this study but the results are unmodified if $p$ is allowed to vary. This is because $p$ is decoupled by construction from the homogeneous EoS parameters, and it additionally does not play any role in the mass fit (see section~\ref{sec:surface}), meaning that it is independent also of the other parameters of the surface tension. The matrix
elements above the diagonal in Fig.~\ref{fig:cormat} give the linear correlation coefficient $r_{ij}=|\sigma_{X_iX_j}|/\sigma_{X_i}\sigma_{X_j}$ obtained for the posterior distribution eq.(\ref{eq:probalikely}) with $w=w_{LD}$, that is after application of the low density filter, while the terms below the diagonal refer to the high density filter. In agreement with the findings of Ref.~\cite{Margueron2018b} we can see that the HD filter does not induce any correlation among the empirical parameters, with the exception of a small positive correlation between $L_{sym}$ and $K_{sym}$, essentially due to the EoS stability requirement at high density. On the other side, many different correlations appear due to the LD constraints. The constraint of mass reproduction induces a clear correlation of the surface tension parameters between themselves, as well as with the zero order isoscalar and isovector parameters ($E_{sat},E_{sym}$), which dominate the global energetics of finite nuclei. More interesting correlations among the different isovector parameters are induced by the constraint of reproducing the ab-initio EFT calculations: besides the well known correlation between $E_{sym}$ and $L_{sym}$ which has been observed by many authors in the context of different models
~\cite{Kortelainen2012,Danielewicz2014,Trippa2008,Colo2014,Holt}
, we also observe a strong correlation among higher order parameters, notably the isovector curvature $K_{sym}$ with the skewness $Q_{sym}$, 
and the high order isoscalar parameters $Q_{sat}, K_{sat}$ with the corresponding isovector ones $Q_{sym}, K_{sym}$. 
These non-trivial correlations can only be observed within the meta-modelling strategy, because in popular functionals like Skyrme the high order parameters are a-priori correlated by the chosen functional form. 
The fourth order parameters $Z_{sat},Z_{sym}$ do not show any correlation with any other parameter, showing their negligble influence on the density relatively close to saturation implied in the LD filter. Finally, the isoscalar effective mass and effective mass splitting are also essentially uncorrelated with the others: variations of these parameters, which play a crucial role in the structure of finite nuclei, are fully compensated by variations of the density derivatives as long as only the total energetics (kinetic plus potential) is involved~\cite{Chatterjee2017}. 

\begin{figure*}[htbp]
    \centering
    \includegraphics[scale=0.45]{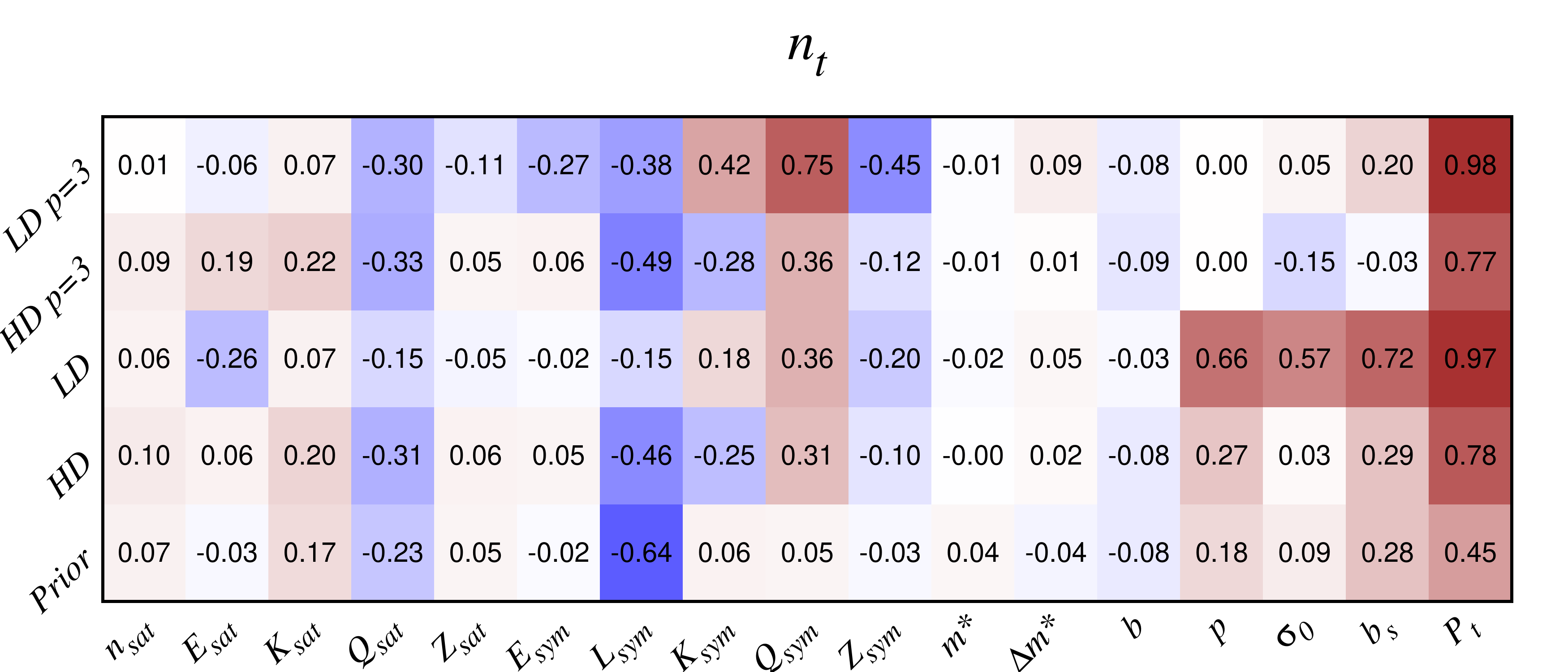}
    \includegraphics[scale=0.45]{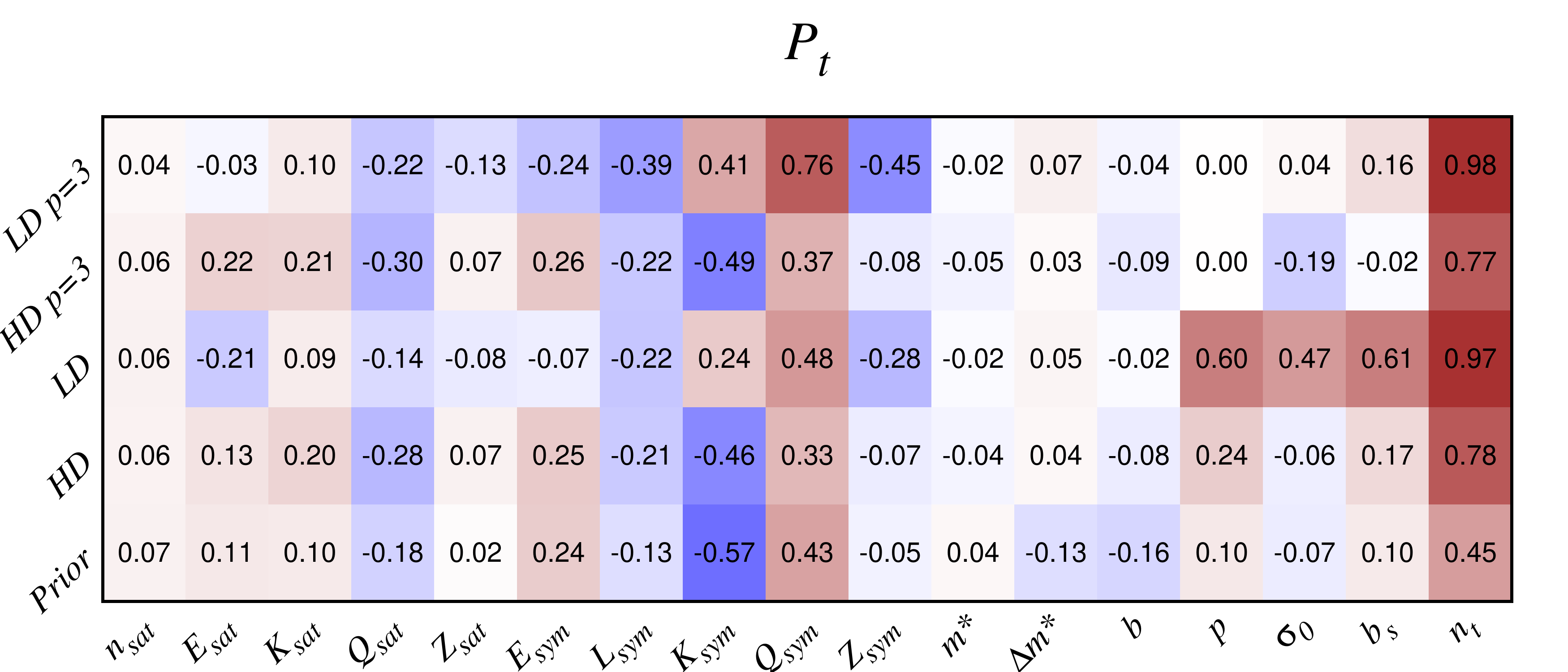}
    \caption{Top: Correlation between the transition density $n_t$ and the parameters for different filters. Bottom: Same for the transition pressure $P_t$.} \label{fig:corntpt}
\end{figure*}

The correlation matrix between the model parameters and the density and pressure of the transition point is presented in Fig.~\ref{fig:corntpt}. 
 When the prior parameter distribution is used, which supposes the EoS parameters fully uncorrelated, the transition density is only (slightly) negatively correlated to the $L_{sym}$ parameter, as it was previously reported~\cite{Ducoin2011}. The transition density directly depends on the energy of $\beta$- equilibrium matter.  The transition pressure being linked to the first derivative of the energy density, it is not surprising that it is correlated to higher order parameters of the symmetry energy, namely $K_{sym}$ and $Q_{sym}$. Since these parameters widely vary in existing functionals, this can explain why the present predictions of the transition pressure are so largely scattered (see Table~\ref{tab:ducoin}).  Once physical correlations among the EoS parameters are accounted for in the posterior distribution, new correlations appear for the transition point.

As we have already discussed, for the computation of the transition point it is important to specify, together with the different parameters analyzed in Fig.~\ref{fig:cormat}, also the isovector surface tension parameter $p$. Let us first consider the case where we fix this parameter to its canonical value $p=3$ that best reproduces the homogeneous matter dynamical spinodal (see Table~\ref{tab:ducoin}). In this case, we can see that the correlation of $n_t$ (resp. $P_t$) with $L_{sym}$ (resp. $K_{sym})$ is preserved.  Further interesting correlations with the high order isovector parameters $K_{sym}$ and $Q_{sym}$ emerge if the EoS sample is restricted to respect the HD filter, and even more if the LD filter is applied.
All these correlations fade away if the $p$ parameter is allowed to vary (lines "LD+masses"). In that case, the transition point is solely correlated to the surface properties (parameters $\sigma_0$, $b_s$ and $p$). This means that the dominant parameter determining the CC transition is the isovector surface tension. 
Only if we constrain its behavior ar large isospin values imposing $p=3$, the correlation with the symmetry energy is recovered. In that case, we can see that the knowledge of the largely discussed $L_{sym}$ parameters is not enough, and higher order parameters beyond $L_{sym}$ must be constrained to improve the prediction of the transition point, both in density and in pressure. 

\begin{figure}[htbp]
    \centering
    \includegraphics[scale=0.47]{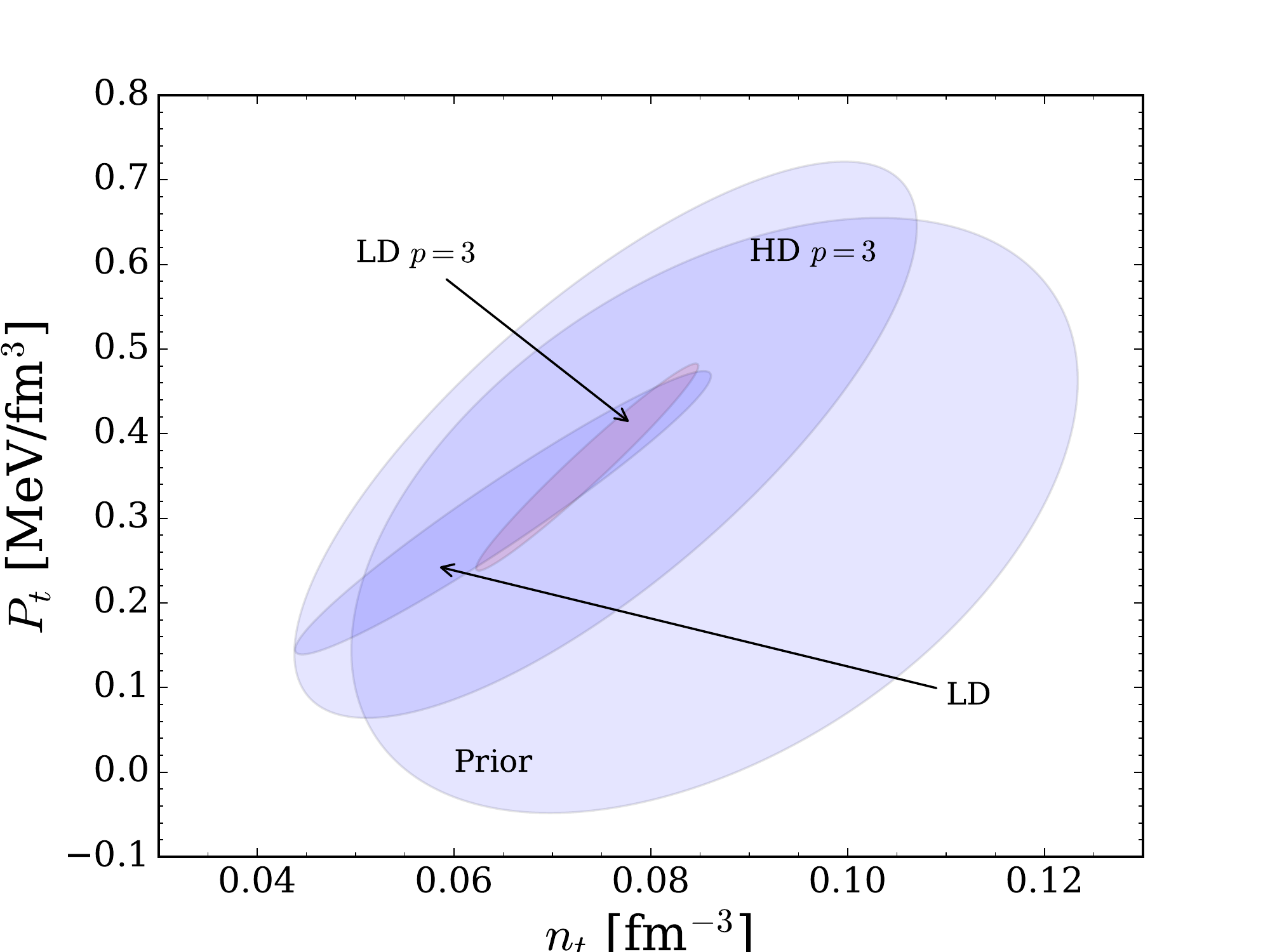}
    \caption{1$\sigma$ error ellipse between the transition density $n_t$ and the transition pressure $P_t$ for different filters.} \label{fig:final}
\end{figure}

Our final result for the value and $1\sigma$ confidence interval for the transition point is displayed in Fig.~\ref{fig:final}. We can observe in a graphical way that the LD and HD filter lead to compatible predictions for the transition point, but the LD filter is by far more constraining.
It is clear from this figure that a reliable determination of the CC transition point demands a better control on the isovector properties of the nuclear surface for extreme isospin values, more than more stringent constraints on the EoS parameters. A Bayesian analysis of the microscopic evaluation of the isovector surface energy within the extended Thomas-Fermi method is under progress. 

\section{Conclusions}\label{sec:conclusions}

In this paper we have presented a detailed study of the core-crust transition point, in the framework of a unified EoS treatment where the inhomogeneous crust is calculated with the same energy functional as employed for the modelling of the homogeneous core. The full parameter space of the EoS, including the successive isoscalar and isovector derivatives of the energy functional at saturation up to $N=4$, the isoscalar and isovector effective mass, and the isospin dependent surface tension, is evenly explored within a completely uncorrelated flat prior, within intervals compatible with the present empirical constraints. The correlated $(2(N+1)+3)$-dimensional parameter distribution is obtained constraining the parameter space such as to reproduce measured mass of magic and semi-magic nuclei, fulfill basic physical conditions as well as modern ab-initio calculations of symmetric and neutron matter.  
We find that the most influential parameter for the determination of the transition point is linked to the behavior of the surface tension for extreme isospin values, specifically the isospin value at which the surface tension vanishes. Only if this parameter is fixed to an educated, but somewhat arbitrary value, important correlations with the isovector parameters of the EoS are recovered. We confirm the correlation of the transition density with the slope of the symmetry energy at saturation $L_{sym}$ already observed in previous works, and additionally point out the important correlation of the transition pressure with the isovector compressibility $K_{sym}$. 

Ab-initio calculations of the symmetric and neutron matter energy and pressure for densities below saturation from Ref.~\cite{Drischler2016} provide very stringent constraints on the EoS parameter distribution, and are shown to be much more effective for the determination of the transition point than the available constraints at supersaturation density. 

Still, if the isovector surface tension is not further constrained, considerable uncertainties affect the transition point ($n_t=0.060\pm 0.027$ fm$^{-3}$, $P_t=0.25\pm 0.225$ MeV/fm$^{-3}$ at the 1$\sigma$ level). If the educated guess $p=3$ can be confirmed by microscopic calculations, the predictions are sensibly changed to $n_t=0.072\pm 0.011$ fm$^{-3}$, $P_t=0.339\pm 0.115$ MeV/fm$^{-3}$. These values, with the associated uncertainty intervals, determine the crustal width and momentum of inertia of neutron star, which in turn allows quantifying the role of the NS crust in interesting astrophysical phenomena such as pulsar glitches.
Our study is performed in the framework of the compressible liquid drop (CLD) model for the inhomogeneous crustal matter and a specific parametrized functional form for the surface energy, but equivalent parameters to our $p$ parameter can be found in alternative modellings, such as for instance isovector gradient couplings in the DFT or Thomas-Fermi approximation~\cite{Bender2011}, or the surface stiffness parameter in the context of the Droplet model~\cite{Warda2009,Centelles2010,Mondal2016}.
 
For this reason, further studies of the isovector surface tension will be extremely important.

 \begin{acknowledgments}
This work was partially supported by the IN2P3 Master Project MAC, "NewCompStar" COST Action MP1304, PHAROS COST Action MP16214.
\end{acknowledgments}

\appendix
\section{More on the properties of EFT EoS}

\begin{figure}[htbp]
    \centering
    \includegraphics[scale=0.39]{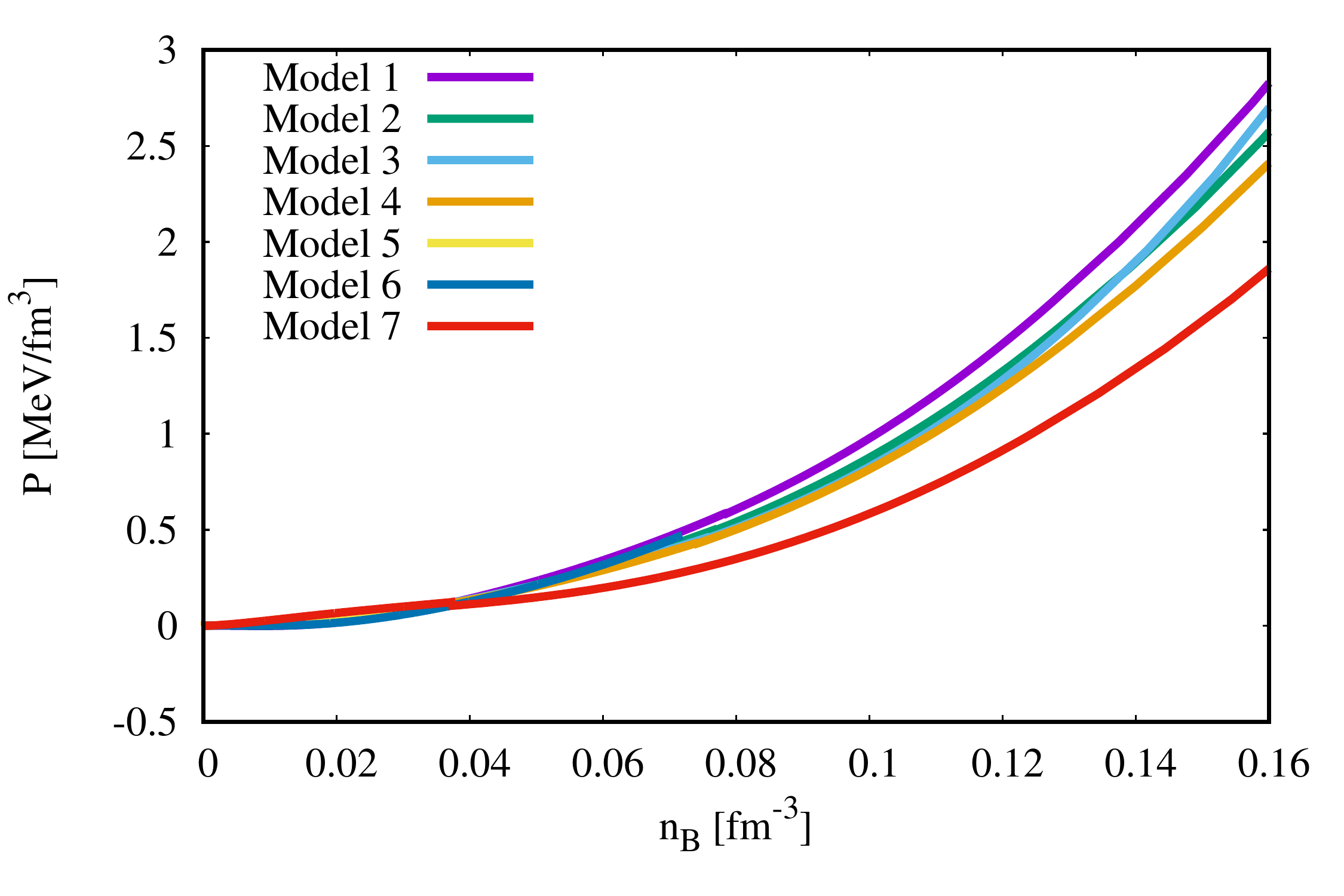}
    \caption{Unified equation of state $P$ as a function of the baryon density $n_B$ (crust and core) for the meta-model realization of the seven Drischler et al. functionals. The empirical parameters for each functional are given in Table IX of Ref.~\cite{Margueron2018a}.}\label{fig:EFT2}
\end{figure}

The EFT constraints displayed in Fig.~\ref{fig:EFT1} are obtained from Ref.~\cite{Drischler2016} as the limiting surface containing the seven different functionals proposed, which correspond to different hypotheses for the NN interaction, resolution scale, cut-offs, and long range couplings. It can also be interesting to compute the transition point using directly the functionals of Ref.~\cite{Drischler2016}. This can be done within our meta-modelling technique considering seven $\vec X$
parameter sets corresponding to the different functionals, which allows computing the energy density for different isospin ratios. The results for the beta-equilibrium EoS and the transition point are displayed in Table~\ref{tab:EFT} and Fig.~\ref{fig:EFT2}.

\begin{table}[htbp]
\centering
\begin{tabular}{ccccc}
\hline\hline
Model & $n_d$ (fm$^{-3}$) & $n_t$ (fm$^{-3}$) & $P_t$ (MeV/fm$^3$) & $n_\mu^*$ (fm$^{-3}$)  \\
\hline
 1 & $2.55\times10^{-4}$ & 0.0785 & 0.5833 & 0.1275 \\
 2 & $2.82\times10^{-4}$ & 0.0769 & 0.4982 & 0.1289 \\
 3 & $2.88\times10^{-4}$ & 0.0758 & 0.4575 & 0.1318 \\
 4 & $2.88\times10^{-4}$ & 0.0740 & 0.4266 & 0.1300 \\
 5 & $1.59\times10^{-4}$ & 0.0320 & 0.0847 &        \\
 6 & $2.82\times10^{-4}$ &        &        &        \\
 7 & $2.68\times10^{-4}$ & 0.0374 & 0.1045 & 0.1244 \\
\hline\hline
\end{tabular}
\caption{Neutron drip density $n_d$, crust-core transition density $n_t$ and pressure $P_t$, and density $n_\mu^*$ at which muons appear for the meta-model realization of the seven Drischler et al. functionals. Empty cells when no solution is found.} \label{tab:EFT}
\end{table}

For some functionals, no solution can be found for the variational equations in the whole density range covered by the crust: this is in particular the case for model 6, which predicts  a very low value of $E_{sat}$ not compatible with values extracted from empirical information from nuclear mass.
Apart from these slight anomalies, the results are well within the range of our predictions imposing the global constraint of Fig.~\ref{fig:EFT1}.

\end{document}